\documentclass[prl,amssymb,floatfix,twocolumn,superscriptaddress,10pt]{revtex4-1}

\usepackage{float}
\usepackage{graphics}

\usepackage{amsmath}
\usepackage{amssymb}
\usepackage{amsthm}
\usepackage{amsfonts}
\usepackage{listings}
\usepackage{enumerate}
\usepackage{latexsym}
\usepackage{psfrag}
\usepackage{bm}
\usepackage[all]{xy}
\usepackage{graphicx}
\usepackage[caption=false]{subfig}
\usepackage{array}
\usepackage[pdftex,colorlinks=true]{hyperref}
\usepackage{color}
\usepackage{mathtools}
\usepackage{tabularx}

\begin{document}
\title{Topological $\mathbb{Z}_2$ Resonating-Valence-Bond Spin Liquid on the Square Lattice}

\author{Ji-Yao Chen}
\affiliation{Laboratoire de Physique Th\'eorique, C.N.R.S. and Universit\'e de Toulouse, 31062 Toulouse, France}

\author{Didier Poilblanc}
\affiliation{Laboratoire de Physique Th\'eorique, C.N.R.S. and Universit\'e de Toulouse, 31062 Toulouse, France}

\date{\today}

\begin{abstract}
 
A one-parameter family of  
long-range resonating valence bond (RVB) state on the square lattice was previously proposed to describe
a critical spin liquid (SL) phase of the spin-$1/2$ frustrated Heisenberg model. We provide evidence that this RVB state in fact also realises a topological (long-range entangled) $\mathbb{Z}_2$ 
SL, 
%in a finite interval of parameter, 
limited by two transitions to critical SL phases.  The topological phase is naturally
connected to the $\mathbb{Z}_2$ gauge symmetry of the local tensor. 
This work shows that, on one hand, spin-$1/2$ topological SL 
with $C_{4v}$ point group symmetry and $SU(2)$ spin rotation symmetry exists on the square lattice 
and, on the other hand, criticality and nonbipartiteness are compatible.
We also point out that, strong similarities between our phase diagram and the ones of classical interacting dimer models
suggest both can be described by similar Kosterlitz-Thouless transitions. 
This scenario is further supported by the analysis of the one-dimensional boundary state. 
\end{abstract}

\maketitle

\emph{Introduction.} 
% High-Tc and the RVB era
In pioneering work~\cite{Anderson1987}, Anderson proposed the resonating valence bond (RVB) state~\cite{Fazekas1973} as the parent Mott insulator
for high-temperature superconductivity~\cite{BednorzMuller1986}. In contrast to magnetic phases, the insulating RVB state hosts pre-existing 
(resonating) singlet pairs of spins $\frac{1}{2}$ - or valence bonds (VB) - which, upon hole doping, give rise to superconducting (coherent) Cooper pairs. 
Originally, the RVB state, in its simplest version only involves resonating singlets build from nearest neighbor (NN) spins $\frac{1}{2}$.
%was also proposed as a non-magnetic ground state of the frustrated triangular Heisenberg 
%antiferromagnet (HAFM)~\cite{Fazekas1973}.
More recently, a generalised RVB state including
longe-range singlet pairs has been introduced to describe the ground state of the spin-1/2 frustrated Heisenberg antiferromagnet on the square lattice~\cite{Wang2013,Poilblanc2017}.  

In recent years, the notion of topological order~\cite{Wen1990,Wen2013} has progressively emerged as a key concept going beyond 
the traditional Ginzburg-Landau paradigm of spontaneous symmetry breaking~\cite{Landau1937, Ginzburg1950}.  It is at the heart of the
excitement for quantum computing as can be conceptually realized in Kitaev's toric code (TC)~\cite{Kitaev2003}. 
Rokhsar-Kivelson (RK) quantum dimer models~\cite{Rokhsar1988,Moessner2002} on the kagome and triangular lattices 
turned out to host  
dimer liquid phases of the same $\mathbb{Z}_2$ (\emph{i.e.} Ising) topological class as the 
TC~\cite{Misguich2002,Ralko2005}. The kagome NN RVB state also
provides a beautiful example - and maybe the simplest possible - of a $\mathbb{Z}_2$ spin liquid (SL)~\cite{Schuch2012,Poilblanc2012, Wildeboer2012, Yao2012}, 
the spin-1/2 $SU(2)$-symmetric analog of the RK dimer liquids.
Topological order is associated to long-range entanglement providing the roots for the emergence of exotic fractionalized bulk excitations. 
%Anyons can have fractional charge and/or fractional statistics. 
%Although TC's anyons are localized,  the more physical
E.g., the kagome NN RVB state hosts mobile spin-1/2 (electric-like) spinon and spinless (magnetic-like) vison excitations~\cite{Poilblanc2012}. 

Strikingly, NN RVB states turn out to have very different infrared (\emph{i.e.} long-distance) properties depending on the 
bipartiteness or non-bipartiteness of the lattice~\cite{Poilblanc2012}. 
For example, in contrast to its analog on the kagome lattice, the NN RVB state on the square lattice exhibits algebraic (dimer-dimer) correlations~\cite{Alet2010,Sandvik2011}. All spin-$\frac{1}{2}$ NN RVB states are in fact closely related to their 
RK dimer-liquid analogs~\cite{Rokhsar1988}.
On the square lattice, a height field representation can be drawn enabling to construct a coarse-grained 
field theory~\cite{Fradkin1990,Fradkin2013} hosting a stable critical Kosterlitz--Thouless (KT) phase.
The non-orthogonality of the valence bond configurations of the NN RVB state does not affect the critical nature of the state but
only modifies the critical exponent~\cite{Alet2010, Sandvik2011, Damle2012}. 
In this work, we show that introducing longe-range bonds into the (square lattice) NN RVB state -- breaking its bipartiteness nature -- 
leads to a rich phase diagram, including a new 
topological $\mathbb{Z}_2$ SL, bounded by two critical KT phases.

\begin{figure}[hb]
\centering 
	\begin{minipage}[c]{0.1\textwidth}
	\centering
	\subfloat[]{\includegraphics[width=14mm, height = 14mm]{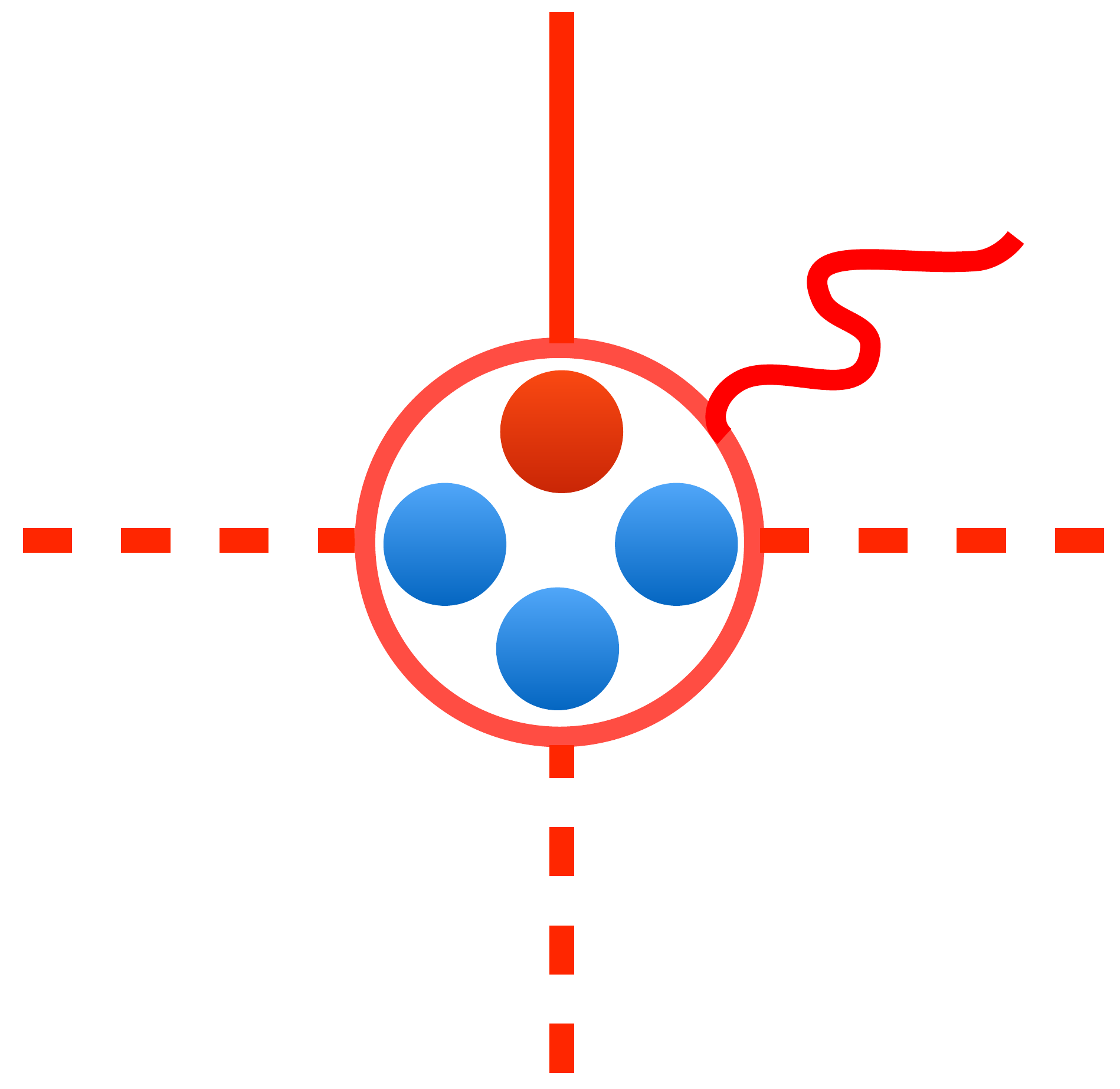}\label{fig:A1_1}}\\
	\subfloat[]{\includegraphics[width=14mm, height = 14mm]{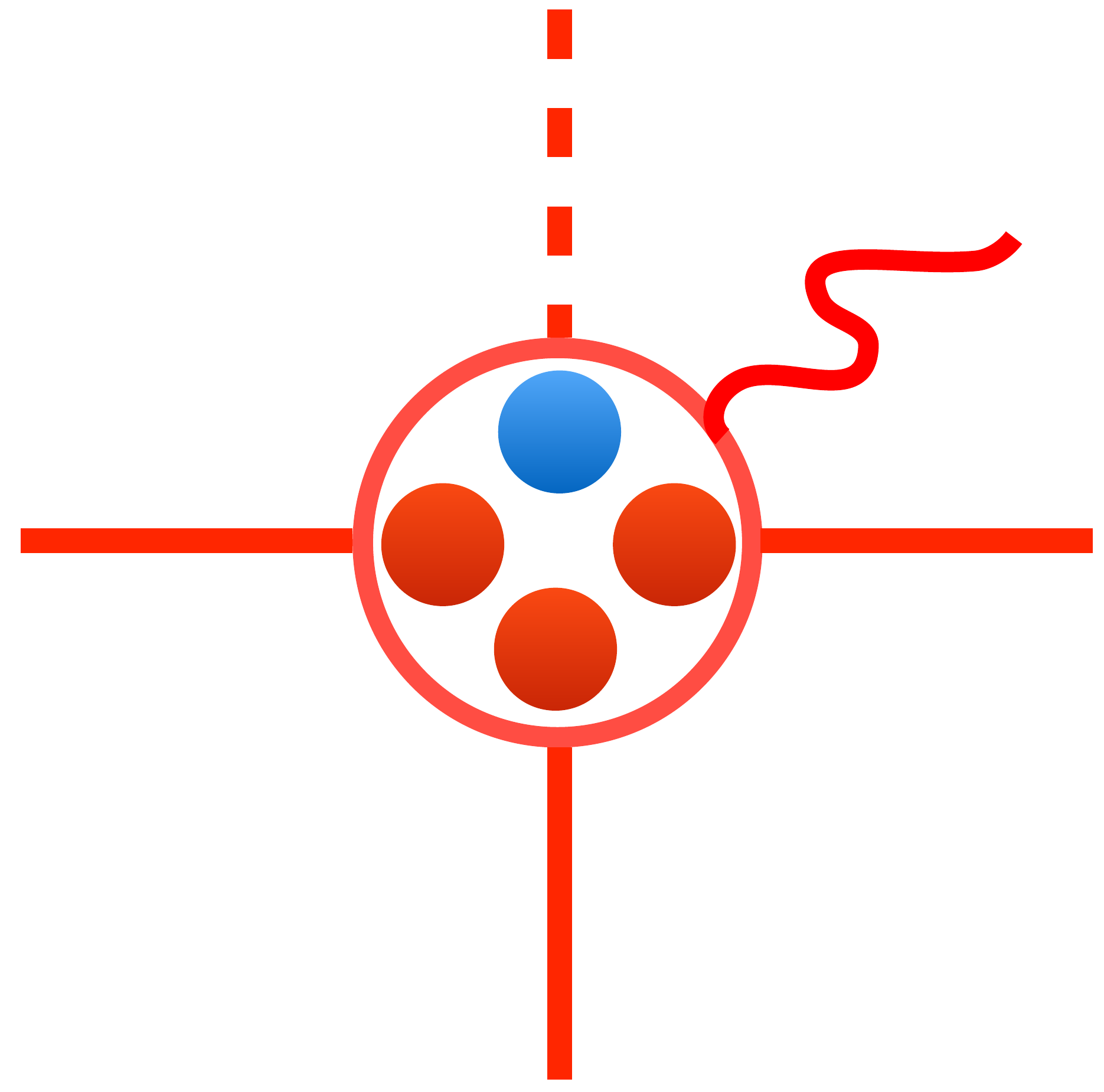}\label{fig:A1_2}}
	\end{minipage}
	\begin{minipage}[c]{0.375\textwidth}
	\centering
	\subfloat[]{\includegraphics[width = 32mm, height = 32mm]{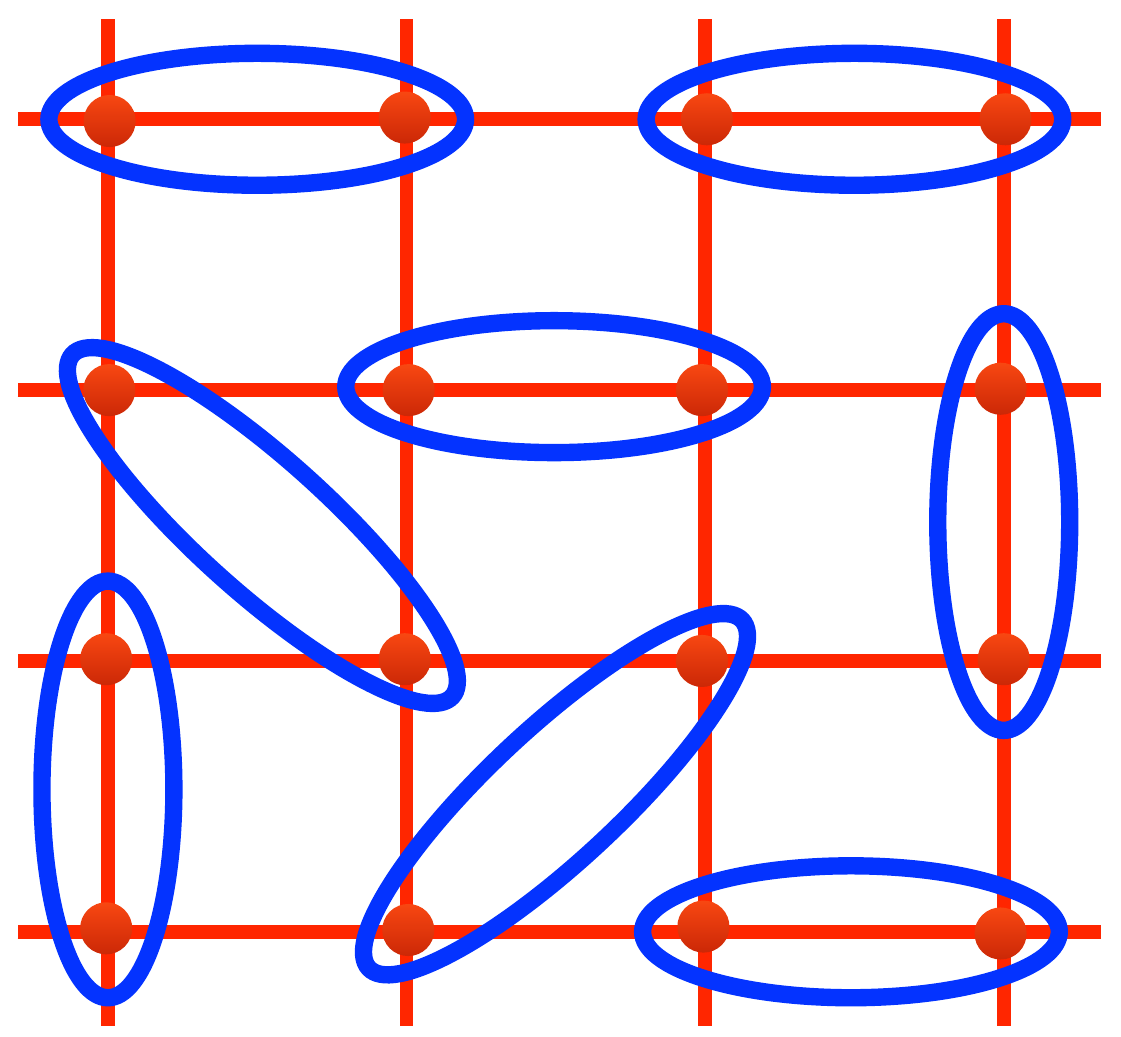}\label{fig:VB}}
	\subfloat[]{\includegraphics[width = 32mm, height = 32mm]{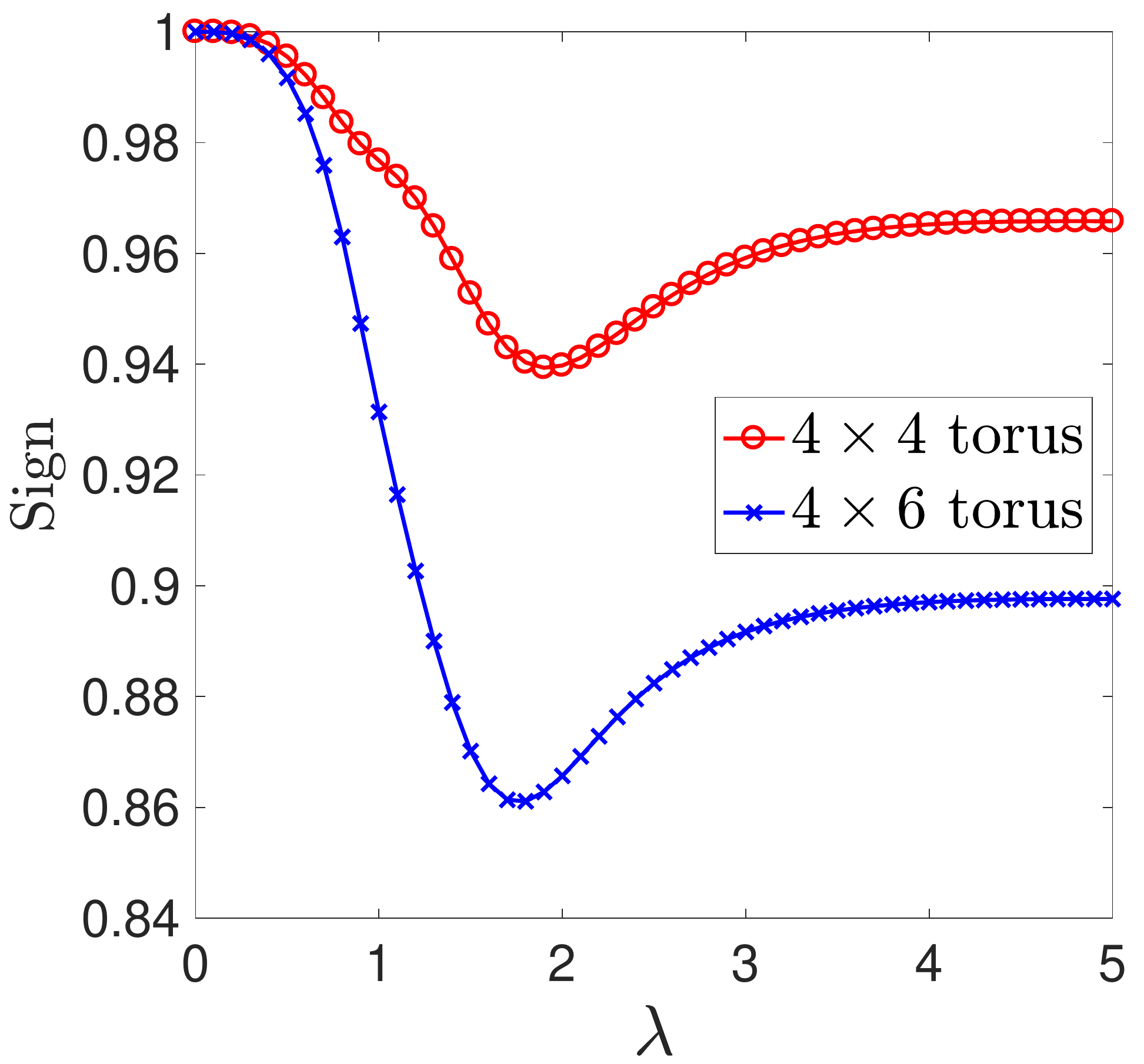}\label{fig:marshall}}\\
	\end{minipage}\\
	\subfloat[]{\includegraphics[width = 78mm, height = 21mm]{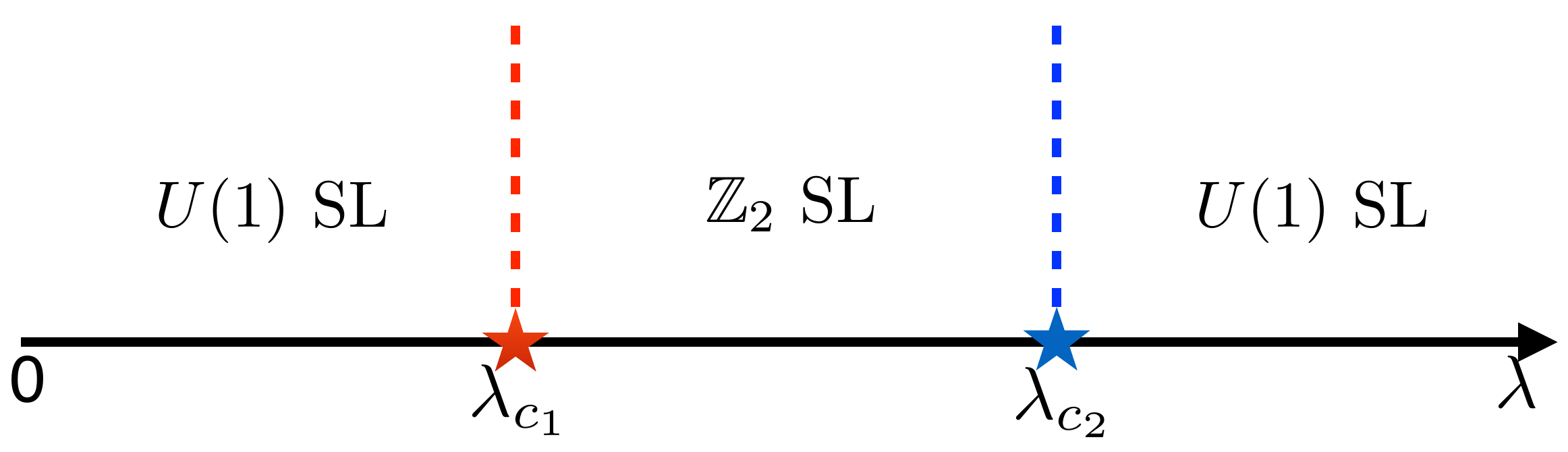}}
\caption{(a,b) Schematic representation of the $\mathcal{A}_1^{(1)}$ and  $\mathcal{A}_1^{(2)}$ site tensors, respectively.
The red (blue) dots represent virtual spins $1/2$ ($0$), solid (dotted) lines represent virtual dimers (absence of dimers) and the red circle stands for the projection operator. The other tensor elements are obtained by rotation and reflection symmetries. (c) Typical VB configuration on the square lattice, 
where ellipses stand for singlet pairs of spin-1/2. (d) Marshall sign versus $\lambda$. 
(e) Phase diagram of the PEPS given by Eq.~(\ref{eq:RVB}) versus $\lambda$.}
\label{fig:schematicFig}
\end{figure}

\emph{The RVB as a PEPS.} For this goal, we consider the generalized RVB state on the square lattice, which was introduced in Ref.~\onlinecite{Wang2013}. 
Such a state is represented as a simple projected entangled-pair state (PEPS) which,
after applying a $\pi$ rotation along $Y$-spin axis on one of the two sublattices, only involves a single tensor $\mathcal{A}$ on every site.
The tensor $\mathcal{A}$ is obtained by
linear combining two $\mathcal{A}_1$ tensors, both of which belong to the  $A_1$ irreducible representation (irrep) of the 
square lattice point group $C_{4v}$:
\begin{equation}\label{eq:RVB}
\mathcal{A} = \mathcal{A}_1^{(1)} + \lambda \mathcal{A}_1^{(2)}.
\end{equation}
The $\mathcal{A}_1^{(1)}$ ($\mathcal{A}_1^{(2)}$) tensor has one (three) virtual spin-$1/2$ and three (one) virtual spin-$0$ in every site configuration and correspondingly one (three) virtual dimer(s) attached to every lattice site.
Virtual spin-1/2 on the bonds connecting NN sites are paired up into singlets.
%These collection of virtual singlets are then entangled by projecting the four virtual spins attached to every site into the physical spin-$1/2$. 
The four virtual spins attached to every site are then projected into the physical spin-$1/2$. 
The bond dimension is thus $D=3$. The elements of $\mathcal{A}_1^{(1,2)}$ which can be found in Ref.~\onlinecite{Mambrini2016} are reproduced in the supplementary materials~\footnote{In the supplementary materials, we list the nonzero elements of the tensors $\mathcal{A}_1^{(1,2)}$, present the CTMRG method for correlation functions and the TRG method for modular matrices. We also define and compute the entanglement entropy of the boundary state, provide the entanglement spectrum of the $\mathbb{Z}_2$ phase~\cite{Cirac2011} 
%and discuss the relevance to frustrated Heisenberg models~\cite{Mambrini2006, Li2017}.} 
and give the expression of the parent Hamiltonian at $\lambda=0$~\cite{Fendley}.} 
for convenience and graphically represented
in Fig.~\ref{fig:schematicFig}(a) and \ref{fig:schematicFig}(b). 
The PEPS formed by $\mathcal{A}_1^{(1)}$ is exactly the bosonic equal weight NN RVB state~\cite{Verstraete2006}, and adding 
the $\mathcal{A}_1^{(2)}$ tensor will generate longer-range VB through quantum teleportation~\cite{Wang2013}.
The violation of the Marshall sign (see Fig.~\ref{fig:schematicFig}(d) and discussion below) implies that singlet VB can appear within the same sublattice, meaning that, strictly speaking, bipartiteness is
broken once $\lambda\neq 0$. A typical VB configuration is shown in Fig.~\ref{fig:schematicFig}(c). Since in every VB configuration the number of $\mathcal{A}_1^{(2)}$ tensor is even, $\mathcal{A}(\lambda)$ and $\mathcal{A}(-\lambda)$ represent the same state and one can restrict to 
\emph{e.g.} $\lambda\ge 0$. 

To draw a phase diagram as a function of the parameter $\lambda$, we have used a corner transfer matrix renormalization group (CTMRG) method~\cite{Nishino1996, Orus2009, Orus2012} taking advantage of the tensor symmetries~\cite{Poilblanc2017}, to compute the spin and dimer correlation functions.
% of the RVB state. 
This has been supplemented by a tensor renormalization group (TRG) analysis to extract topological properties,
% of the RVB state, 
if any. Our results are summarized in the schematic phase diagram shown in Fig.~\ref{fig:schematicFig}(e).  
A short-range topological $\mathbb{Z}_2$ SL phase is found in an extended region $\lambda\in (\lambda_{c_1},\lambda_{c_2})$, surrounded by two critical SL phases, 
where $\lambda_{c_1}=0.85(5)$, $\lambda_{c_2}=2.85(5)$. 
We emphasize the existence of
an emergent U(1) gauge field responsible for the critical nature of the SL phases at $\lambda<\lambda_{c1}$ and $\lambda>\lambda_{c2}$. Next, we present our numerical results supporting this phase diagram.  

\emph{Marshall sign and gauge symmetry.} 
The PEPS we are considering is a $SU(2)$ spin singlet and can be expressed as coherent superposition of valence bond configurations, which form an overcomplete basis:
\begin{equation}\label{eq:VB}
|\psi(\mathcal{A})\rangle = \sum c_{(i_1j_1),(i_2j_2),...}|(i_1j_1),(i_2j_2),... \rangle,
\end{equation}
where $|(i_1j_1),(i_2j_2),... \rangle$ is a VB configuration and $c_{(i_1j_1),(i_2j_2),...}$ is the corresponding amplitude.
Note that, in general we cannot factorize $c_{(i_1j_1),(i_2j_2),...}$ as a product of weights function of the dimer length.
 A central question to ask is what is the $(ij)$ singlet pairing type in the VB basis, \emph{i.e.}, whether there is only inter-sublattice $AB$ pairing. To answer this question, we have investigated the Marshall sign~\cite{Marshall1955} in the Ising basis. We put the PEPS on a finite lattice with torus geometry, and use exact contraction to obtain the wave function. Then we compute the Marshall sign average, defined as $\langle \mathrm{sign} \rangle = \frac{\sum_{c}\mathrm{sgn}_c|\langle c|\psi(\mathcal{A}) \rangle|^2}{\sum_c |\langle c|\psi(\mathcal{A})\rangle|^2}$, where $c$ is the Ising configuration and $\mathrm{sgn}_c$ is determined by the sign of the coefficient. As can be seen in Fig.~\ref{fig:schematicFig}(d), 
for arbitrary small $\lambda\ne 0$, the Marshall sign average deviates from 1, and more severely  with increasing system size. 
These results imply that our RVB PEPS cannot be written in the canonical Liang-Dou{\c c}ot-Anderson form~\cite{Liang1988} with only $(AB)$
singlet pairs. Reversely, if VB are present on the same $A$ or $B$ sublattices, it implies that, effectively, the bipartiteness of the lattice is broken. This property is in fact connected to the broken $U(1)$ gauge symmetry of the site tensor: 
for $\lambda=0$  ($\lambda=\infty$) the number of virtual spin-1/2 around each site is fixed to $1$ ($3$), while for $0<\lambda<\infty$, only the {\it parity} of this number is conserved so that the U(1) gauge symmetry is broken down 
to $\mathbb{Z}_2$.  The two $U(1)$-symmetric RVB states are in fact closely related to their RK critical
 dimer-liquid analogs~\cite{Rokhsar1988} 
for which the KT algebraic (dimer) correlations follow from a coarse-grained height-field theory~\cite{Fradkin1990, Fradkin2013}.
At finite $\lambda$, the long distance field theory can no longer be obtained by the same coarse-graining procedure.  Therefore, it is not clear whether each algebraic phase will survive 
in a finite region of the parameter $\lambda$. Next, we provide numerical evidence for the stability of both 
critical phases and the emergence of a novel short-range SL in between.

\begin{figure}
\centering
	\subfloat[]{
	\includegraphics[width=41.8mm,height=45mm]{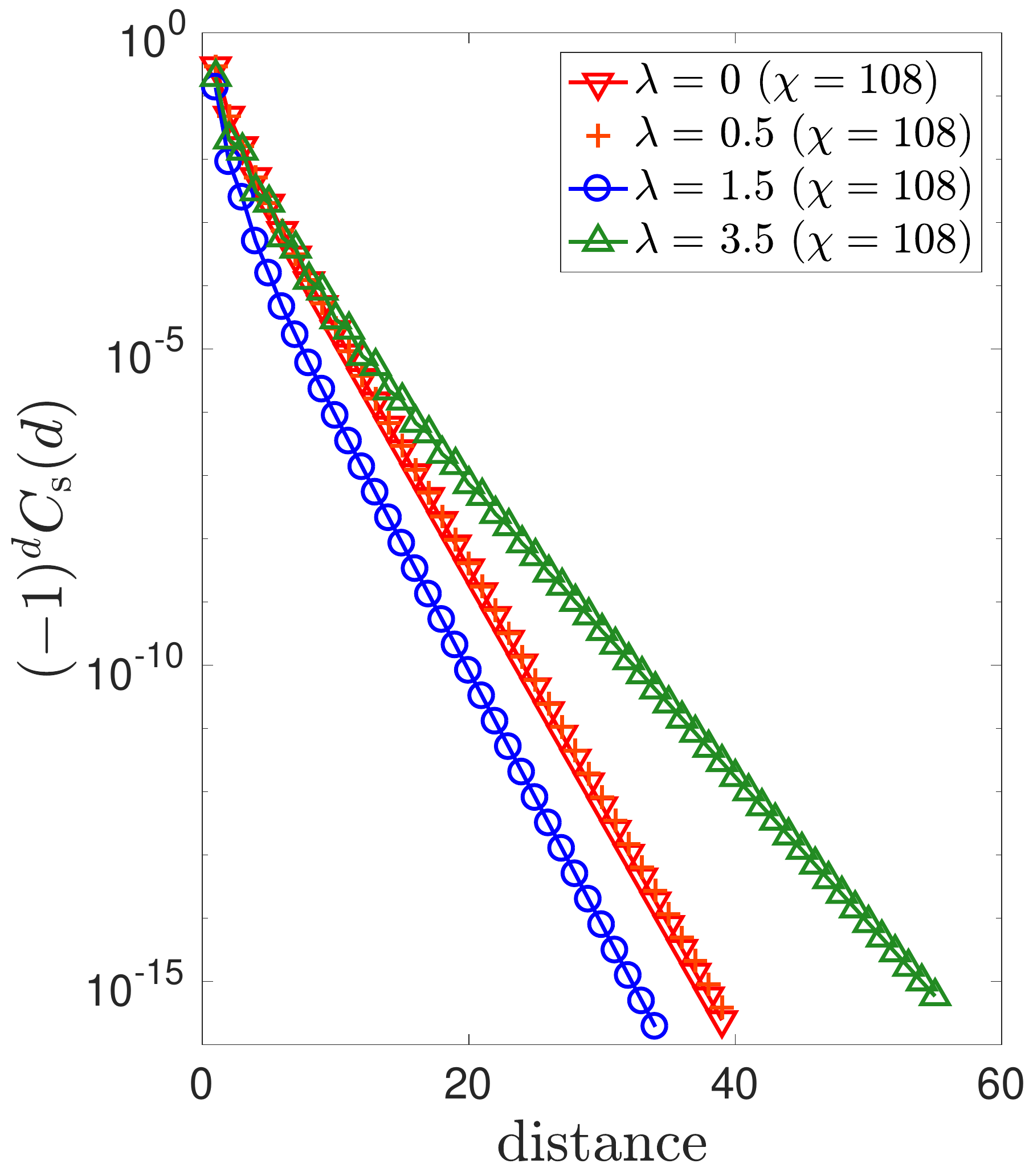}}
	\subfloat[]{
	\includegraphics[width=42mm,height=44mm]{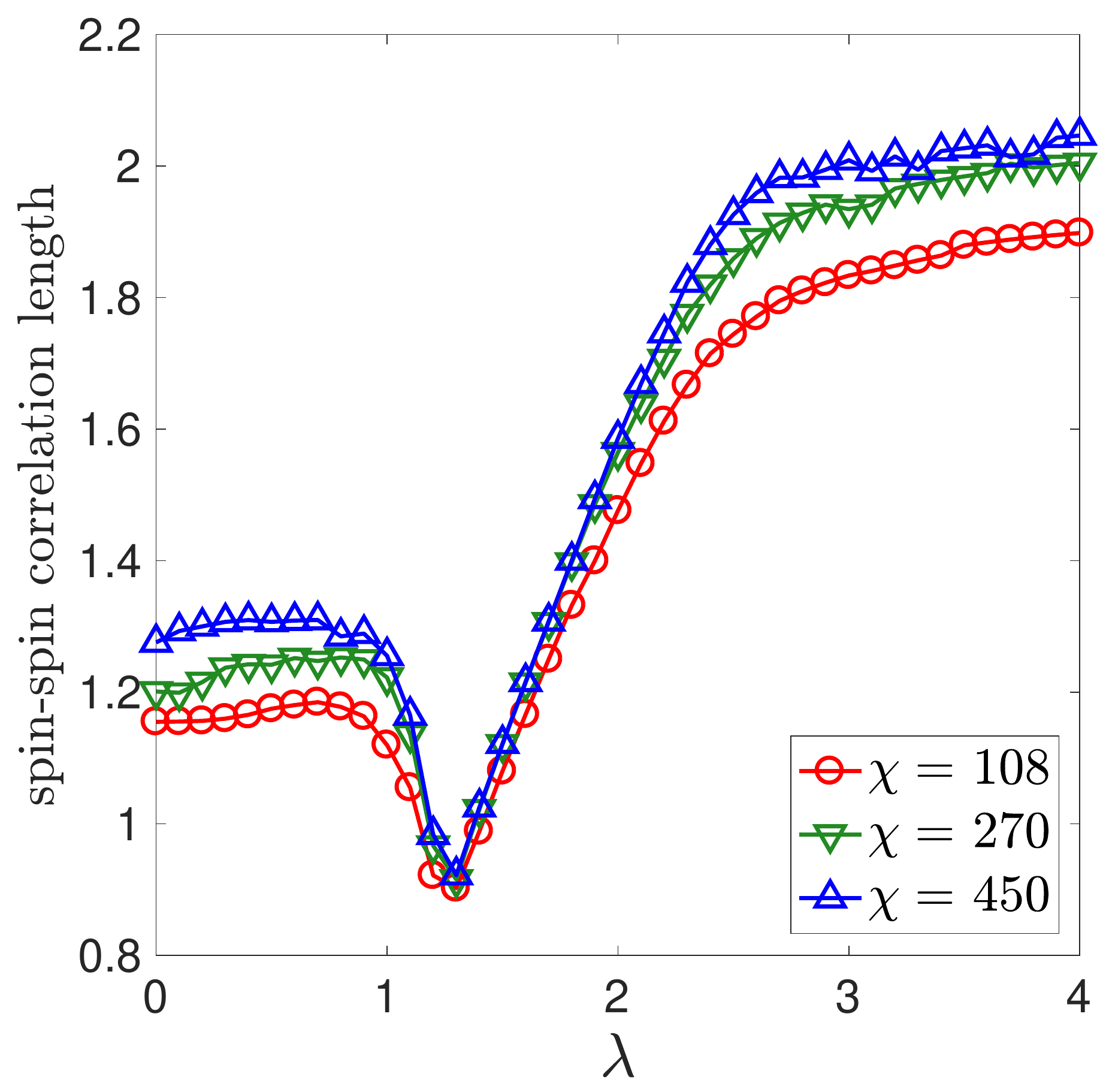}}\\
	\subfloat[]{
	\includegraphics[width=43.5mm,height=43.5mm]{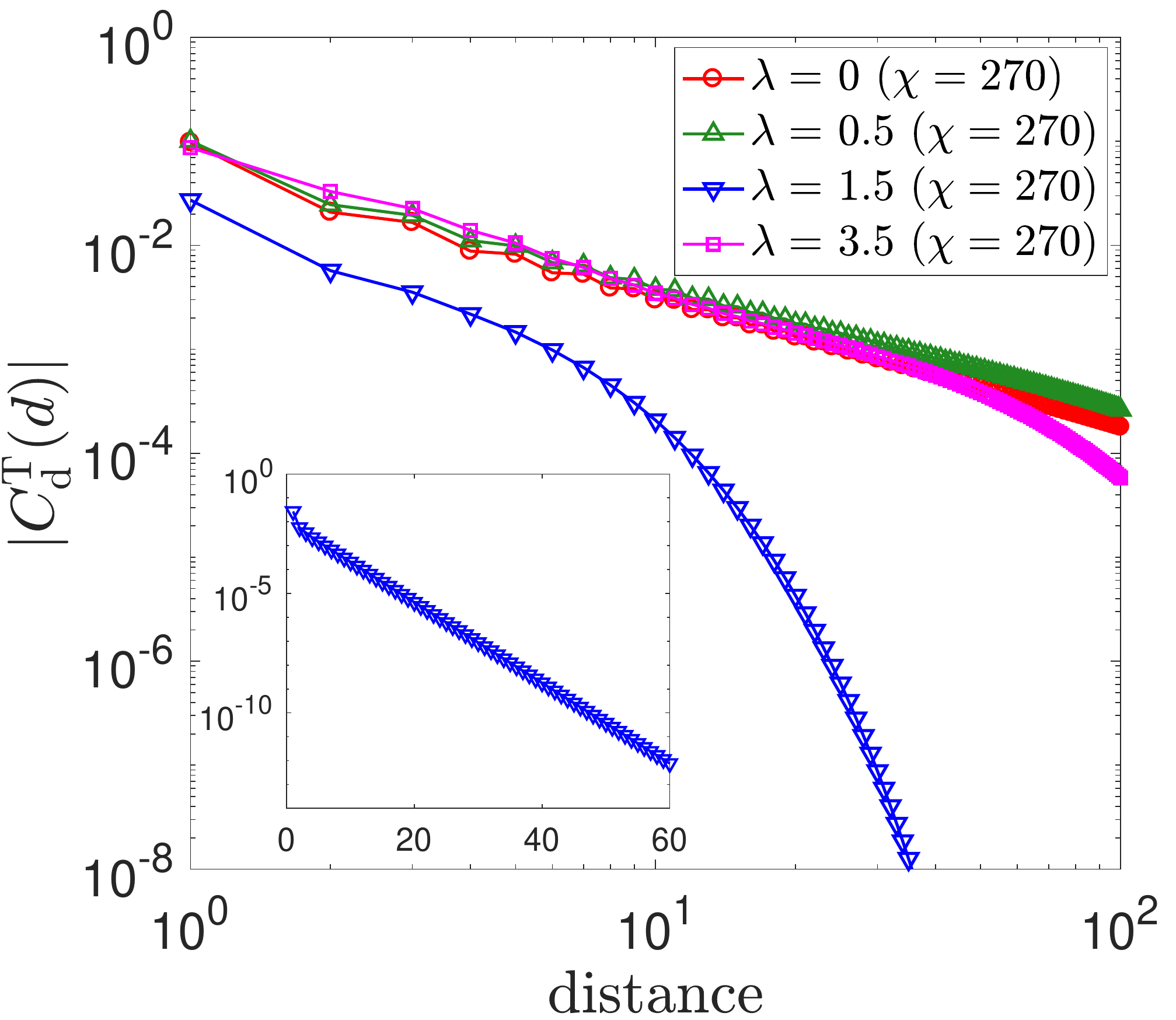}}
	\subfloat[]{
	\includegraphics[width=42mm,height=43.5mm]{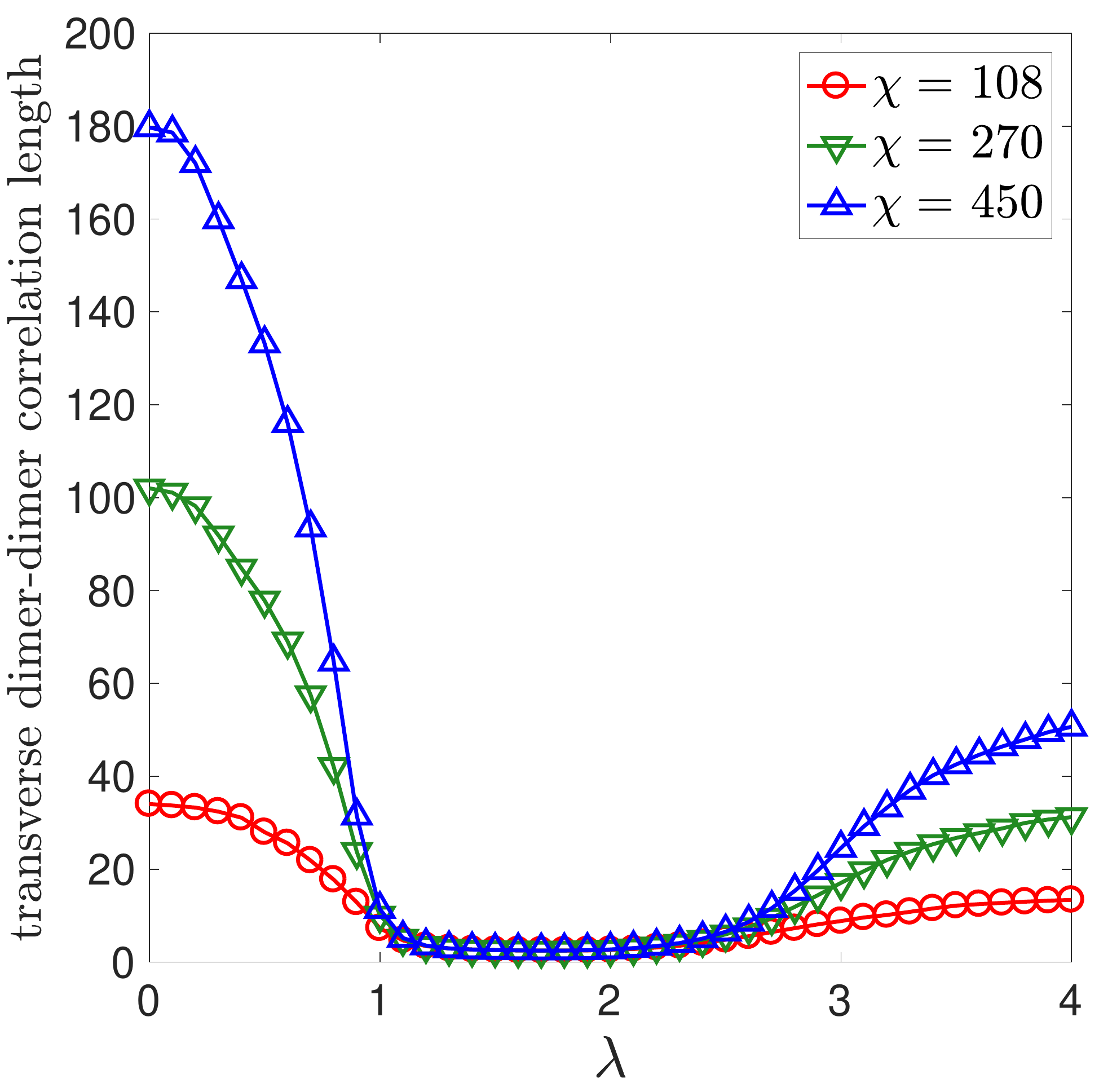}}\\
\caption{(a) Spin correlation function vs distance for different $\lambda$ at fixed $\chi=12D^2$ (semi-log plot).
(b) Spin correlation length vs $\lambda$ for different values of $\chi$. 
(c) Log-log plot of the (transverse) dimer correlation vs distance 
for different $\lambda$ at fixed $\chi=30D^2$ (semi-log plot for $\lambda=1.5$ in inset).
(d) Dimer correlation length vs $\lambda$ for different values of $\chi$ (same as in (b)).
In (b) and (d) the error bar of fitting correlation length is smaller than symbol size and has been omitted.}
\label{fig:corrFunc}
\end{figure}

\emph{Correlation functions.} To calculate the physical observables of the PEPS given by Eq.~\eqref{eq:RVB}, we use the CTMRG method to extract various correlation functions~\cite{Nishino1996, Orus2009, Orus2012, Poilblanc2017}. The CTMRG method allows us to work directly in the thermodynamic limit, whose accuracy is controlled by the bond dimension of environment tensors, denoted as $\chi$. For completeness, we include the details of the specific CTMRG method we are using  in the appendix~\cite{Note1}.
We are interested in spin-spin and longitudinal/transverse dimer-dimer correlation functions along e.g. the $\mathbf{e}_x$ (horizontal) direction defined as:
\begin{equation}\label{spin_spin}
\begin{split}
& C_\mathrm{s}(d) = \langle \mathbf{S_i} \cdot \mathbf{S}_{\mathbf{i}+d\mathbf{e}_x}\rangle_0,\\
& C_\mathrm{d}^{(\mathrm{L})}(d) = \langle D_\mathbf{i}^x D_{\mathbf{i}+d\mathbf{e}_x}^x\rangle_0 - \langle D_\mathbf{i}^x\rangle_0 \langle D_{\mathbf{i}+d\mathbf{e}_x}^x\rangle_0,\\
& C_\mathrm{d}^{(\mathrm{T})}(d) = \langle D_\mathbf{i}^y D_{\mathbf{i}+d\mathbf{e}_x}^y\rangle_0 - \langle D_\mathbf{i}^y\rangle_0 \langle D_{\mathbf{i}+d\mathbf{e}_x}^y\rangle_0.
\end{split}
\end{equation}
where the dimer operators $D_\mathbf{i}^x = \mathbf{S_i}\cdot \mathbf{S}_{\mathbf{i+e}_x}$ and $D_\mathbf{i}^y = \mathbf{S_i}\cdot \mathbf{S}_{\mathbf{i+e}_y}$.
Note that the correlation along the $\mathbf{e}_y$ (vertical) direction are the same due to $C_{4v}$ lattice symmetry.
Also, since L and T dimer correlations give similar results we shall only show the T correlations, for conciseness.

The spin-spin correlations show clear exponentially decay with momentum $(\pi,\pi)$ in all 
parameter region. Typical behaviors
for $\chi=12D^2$ are shown in Fig.~\ref{fig:corrFunc}(a).
By fitting the asymptotic linear behaviors of the data according to $\mathrm{ln}|C_\mathrm{s}(d)| = -(1/\xi_\mathrm{s})d+c_0$, we straightforwardly get the correlation length $\xi_\mathrm{s}$ from the slopes $-1/\xi_\mathrm{s}$, which is shown in Fig.~\ref{fig:corrFunc}(b). It can be seen that the spin-spin correlation length is very short in the full parameter region and, with increasing $\chi$, converges to a small finite value. Note however that a small singularity may be present around $\lambda\simeq 0.9$, reflecting some transition (see next). 

\begin{figure}
\centering
	\subfloat[]{
	\includegraphics[width=42mm,height=43mm]{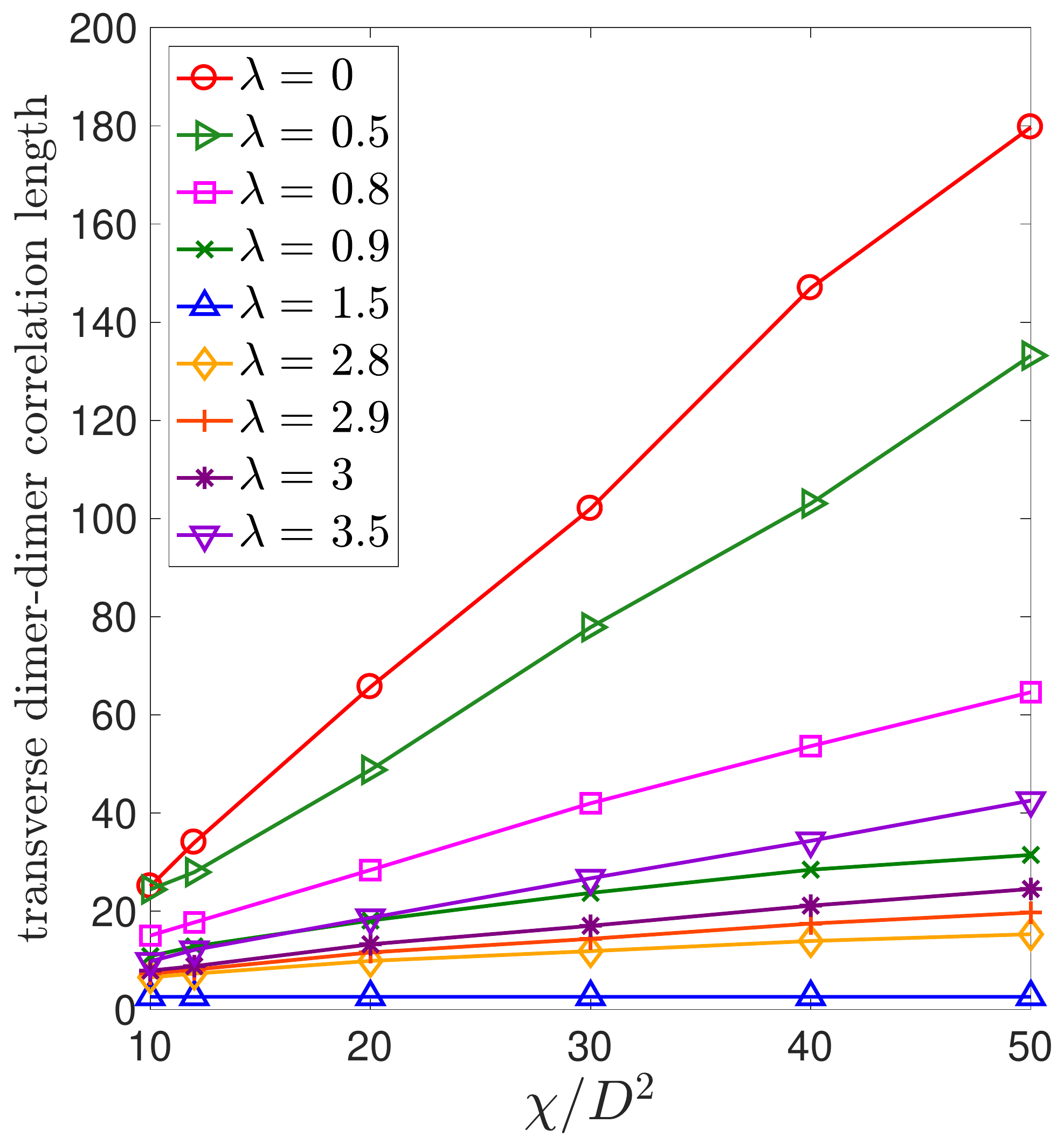}}
	\subfloat[]{
	\includegraphics[width=42mm,height=43mm]{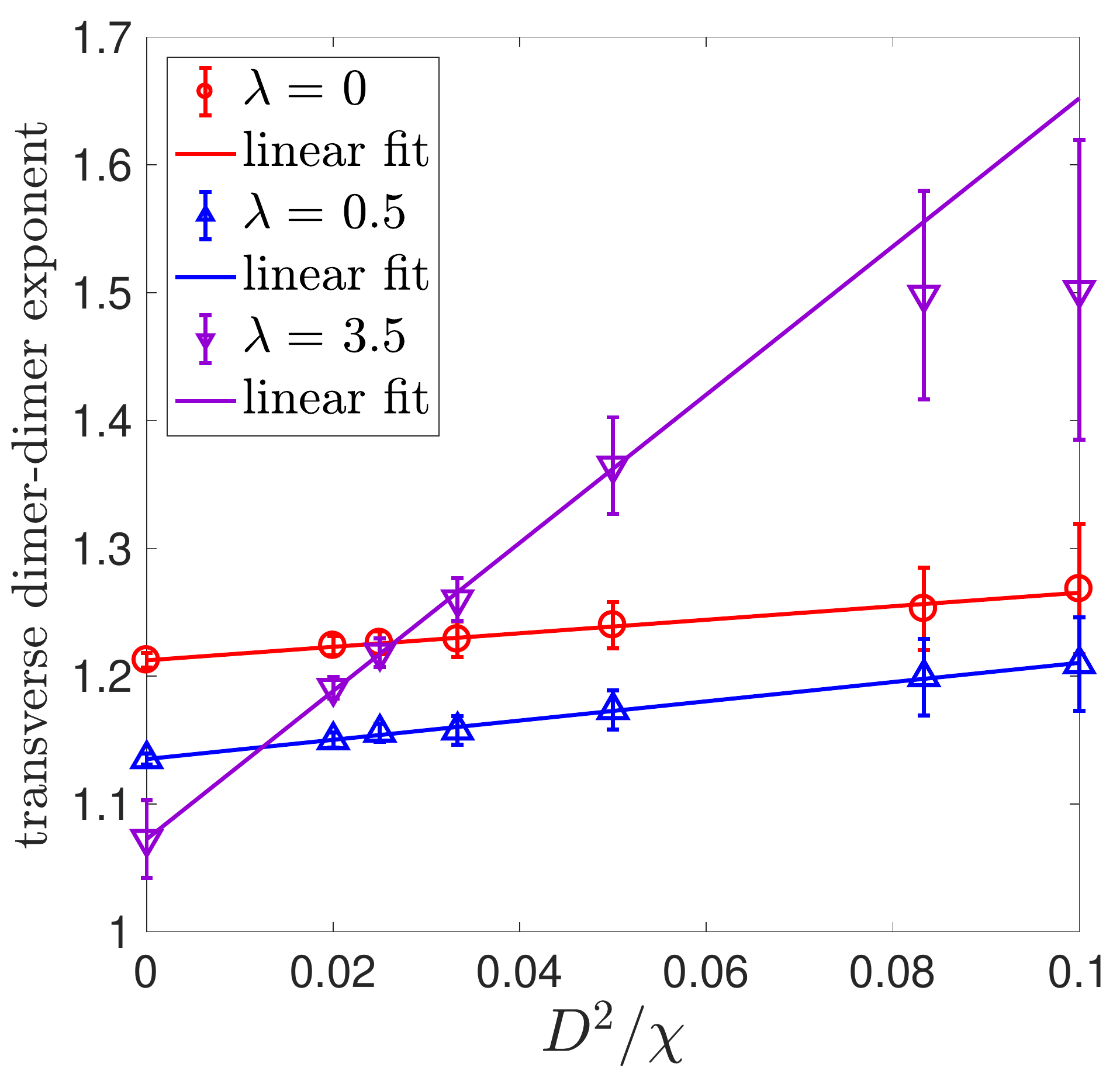}}\\
	\subfloat[]{	
	\includegraphics[width=42mm,height=42mm]{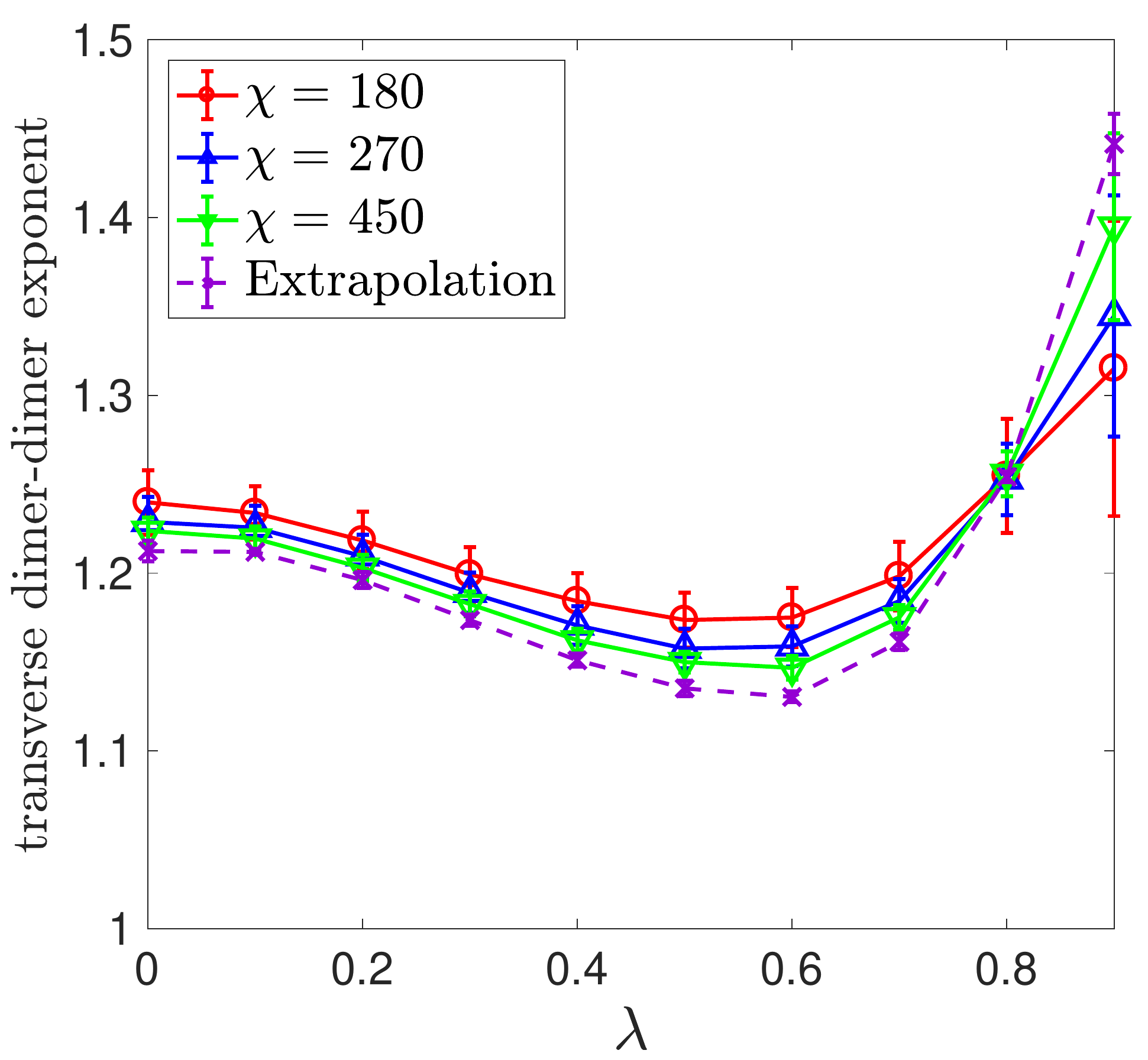}}
	\subfloat[]{
	\includegraphics[width=42mm,height=42mm]{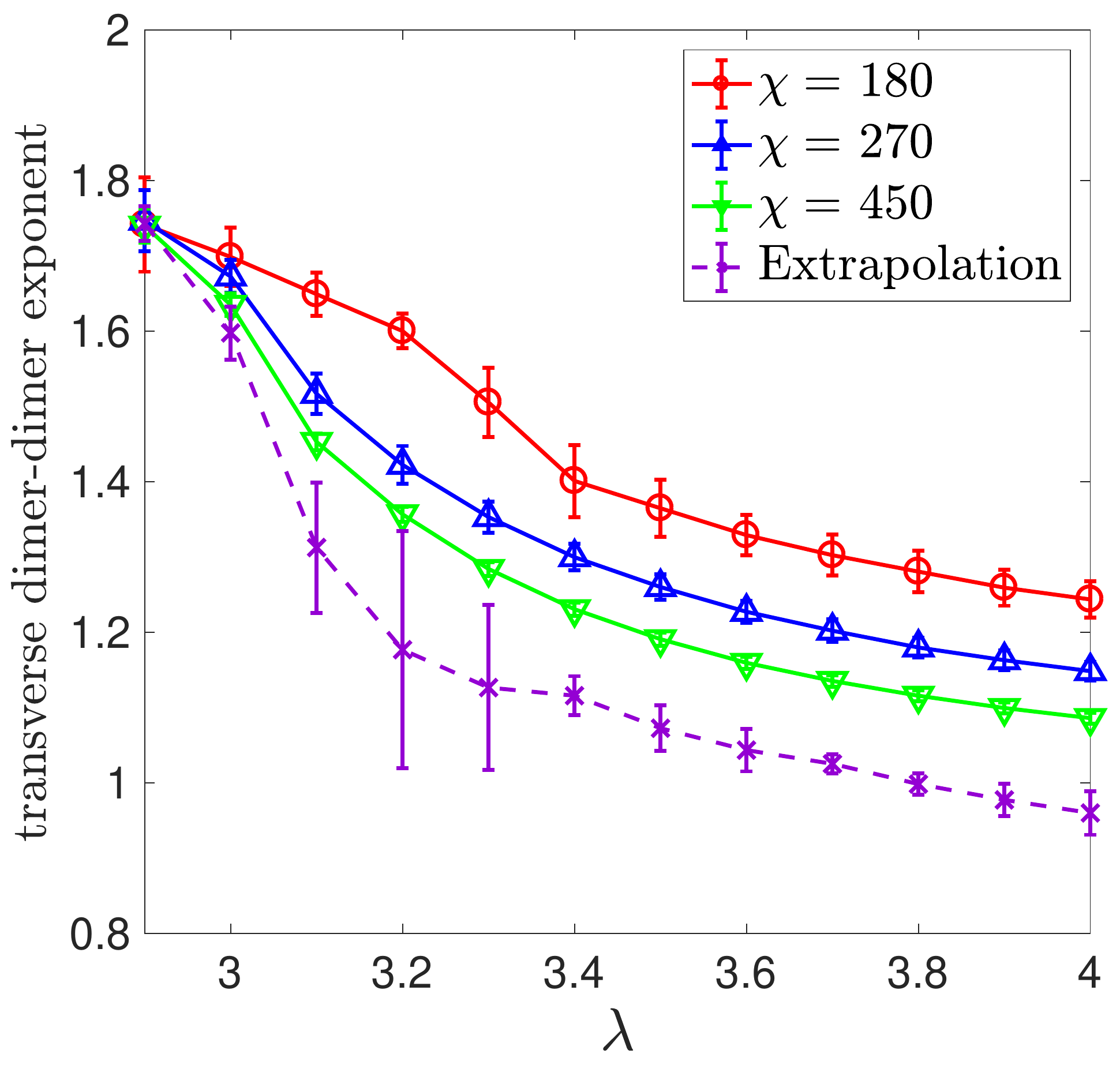}}\\
\caption{(a) Dimer (transverse) correlation length vs $\chi/D^2$ for different values of
$\lambda$. (b) Fits vs $D^2/\chi$ of the exponent of the power-law dimer correlation, for
different values of $\lambda$ in the critical phases. (c,d) Dimer exponent vs $\lambda$ for different values
of $\chi$ and $\chi\rightarrow\infty$ extrapolation. In (b,c,d), the error bar of the exponent at finite $\chi$ comes from 
fitting the dimer correlation functions vs distance, while error bar of extrapolated exponent comes from linear fitting vs $D^2/\chi$.}
\label{fig:exponent}
\end{figure}

The dimer-dimer correlations reveal new exotic features. For both small $\lambda$ ($\lambda<\lambda_{c_{1}}$) and large $\lambda$ ($\lambda>\lambda_{c_{2}}$), the analysis of the data shows clear power-law decaying dimer-dimer correlations as can be seen \emph{e.g.} 
in Fig.~\ref{fig:corrFunc}(c). 
Although for any finite $\chi$, the asymptotic long distance dimer-dimer correlations decay always exponentially, the correlation length $\xi_\mathrm{d}$ (see Fig.~\ref{fig:corrFunc}(d)) never saturates with increasing $\chi$ as can be seen in Fig.~\ref{fig:exponent}(a), which indicates that the two regions are in fact critical. By fitting the critical behavior $|C_d(d)|\sim d^{-\eta}$ in the $d\le \xi_{\mathrm{d}}$ region, we can obtain the critical exponent $\eta$, shown in Figs.~\ref{fig:exponent}(b-d). 
The converged exponent at $\lambda=0$ (NN RVB state) agrees very well with Monte Carlo results~\cite{Alet2010, Sandvik2011}. 
By analysing the behavior of the dimer correlation length 
%and exponent 
with increasing $\chi$, we have located the phase boundaries $\lambda_{c_1}=0.85(5), \lambda_{c_2}=2.85(5)$.
%The critical behavior is also in agreement with the quantum dimer model 
%at the RK point~\cite{Rokhsar1988} and the associated field theory analysis\cite{Fradkin1990, Fradkin2013}. 
Most strikingly, in the 
intermediate $\lambda_{c_1}<\lambda<\lambda_{c_2}$ region, $\xi_{\mathrm{d}}(\chi)$  clearly saturates to a small value, as shown in Fig.~\ref{fig:corrFunc}(d), revealing a true short-range behavior.

\begin{figure}[hb]
\centering
	\subfloat[]{
	\includegraphics[width = 40mm, height = 41mm]{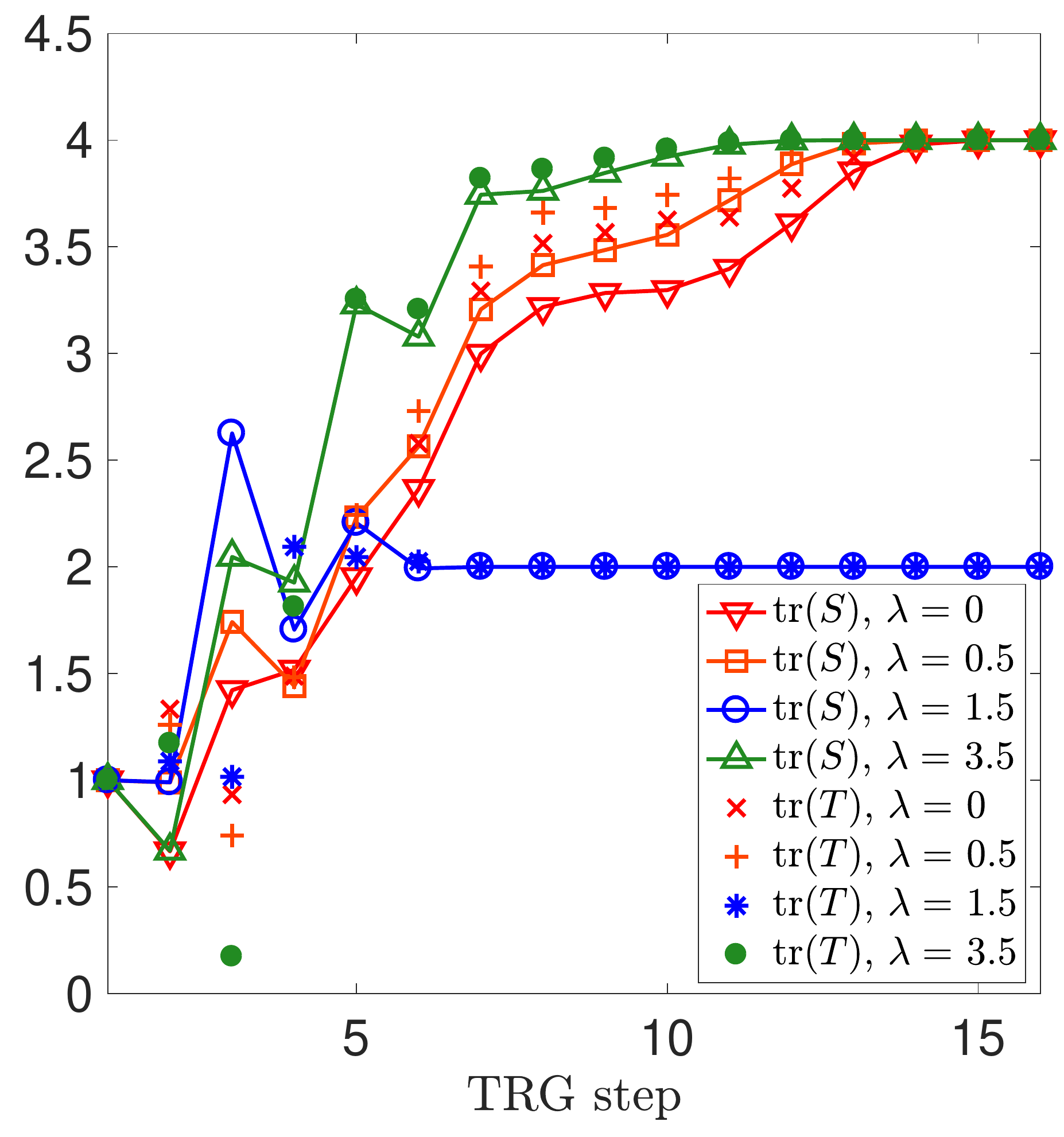}}
	\subfloat[]{
	\includegraphics[width = 42mm, height = 41mm]{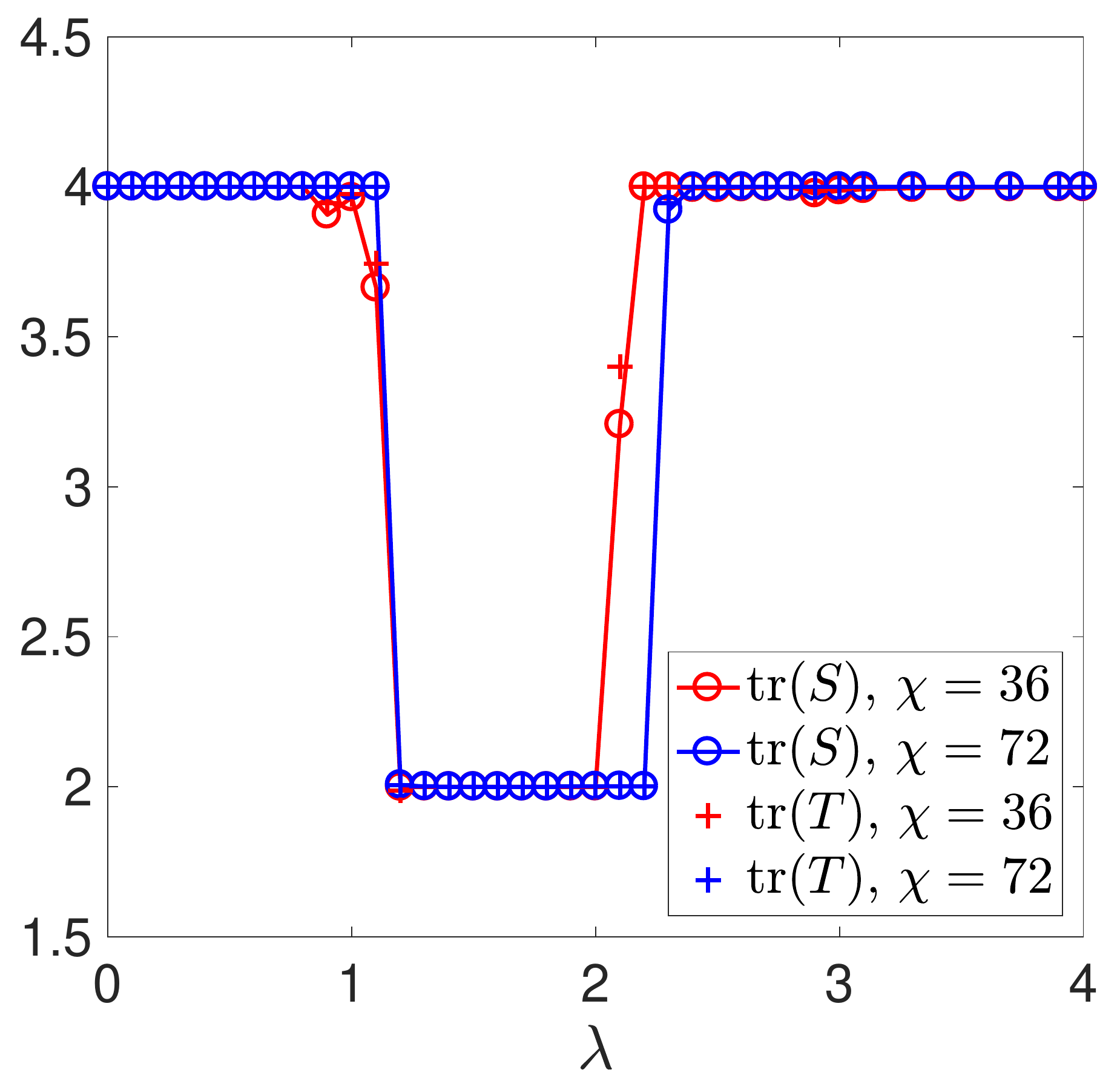}}
\caption{Trace of the modular matrices $S$ and $T$. (a) vs TRG step number at $\chi=8D^2$; (b) vs $\lambda$ after
$12$ and $16$ steps at $\chi=4D^2$ and $8D^2$, respectively.}
\label{fig:ST}
\end{figure}

{\it Search for topological order.} 
The exponentially decaying spin and dimer correlation functions (with extremely short correlation lengths) strongly
support the existence of a new quantum phase between $\lambda_{c_1}$ and $\lambda_{c_2}$. Further more, since
there is no evidence for any symmetry breaking order, it should be a short range spin liquid. Then, a natural question is whether
this spin liquid exhibit topological order.
The PEPS in Eq.~(\ref{eq:RVB}) bears $\mathbb{Z}_2$ gauge symmetry, except at $\lambda=0, \infty$ where higher $U(1)$ gauge symmetry is present. The $\mathbb{Z}_2$ gauge symmetry is generated by $2\pi$ spin rotation, which only induces a minus sign in the $\mathcal{A}$ tensor. We then expect 
$\mathbb{Z}_2$ topological order in the intermediate region. To verify this, we use the TRG method to obtain the modular matrices. Notice that, in order to correctly implement this method, we need to keep the $\mathbb{Z}_2$ gauge symmetry~\cite{He2014, Mei2017}. The TRG method for modular matrices is briefly reviewed in the supplemental materials~\cite{Note1}, whose precision is controlled by the bond dimension $\chi$ of the double tensor. 
After every TRG step, we put the double %${\cal A}\otimes{\cal A}$
 tensor on a torus. Inserting gauge symmetry transformation, we obtain the complete modular $S$ and $T$ matrices. In the intermediate region, 
the modular matrices converge after 6 TRG steps, while it takes much longer (typically 10--12 steps) to obtain converged results in the critical regions, as shown in Fig.~\ref{fig:ST}(a). 

The converged modular matrices for the short range SL are:
\begin{equation}\label{eq:Z2ST}
S=\left(\begin{smallmatrix}1&0&0&0\\0&0&1&0\\0&1&0&0\\0&0&0&1\end{smallmatrix}\right),	\quad T=\left(\begin{smallmatrix}1&0&0&0\\0&1&0&0\\0&0&0&1\\0&0&1&0\end{smallmatrix}\right),
\end{equation} 
which are identical to the modular matrices of the TC in the string basis.
For the two critical regions, we also obtain converged modular matrices:
\begin{equation}\label{eq:trivialST}
S=\left(\begin{smallmatrix}1&1&1&1\\1&1&1&1\\1&1&1&1\\1&1&1&1\end{smallmatrix}\right),	\quad T=\left(\begin{smallmatrix}1&1&1&1\\1&1&1&1\\1&1&1&1\\1&1&1&1\end{smallmatrix}\right),
\end{equation} 
which are rank-1 matrices and indicate trivial topological order.
The trace of the converged modular matrices for different $\lambda$ is shown in Fig.~\ref{fig:ST}(b), where sharp transitions can be seen between the different regions. 
We note that, similar topological information can also be obtained by investigating the leading eigenvalues of the transfer matrix in different topological sectors when putting the PEPS on an infinitely long cylinder~\cite{Schuch2013}.

\begin{figure}[hb]
\centering
	\subfloat[]{
	\includegraphics[width = 40mm, height = 39mm]{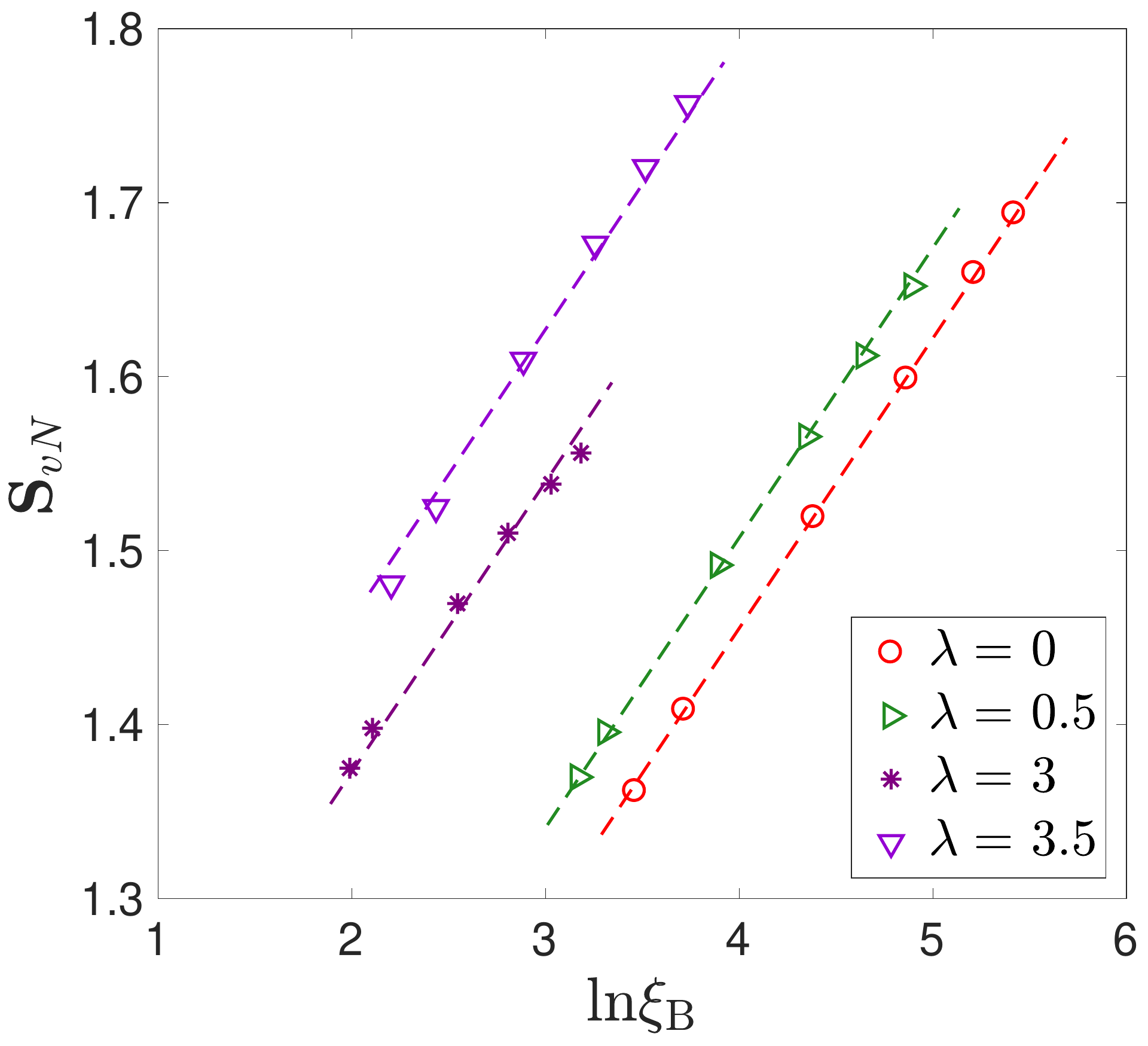}}
	\subfloat[]{
	\includegraphics[width = 40mm, height = 39mm]{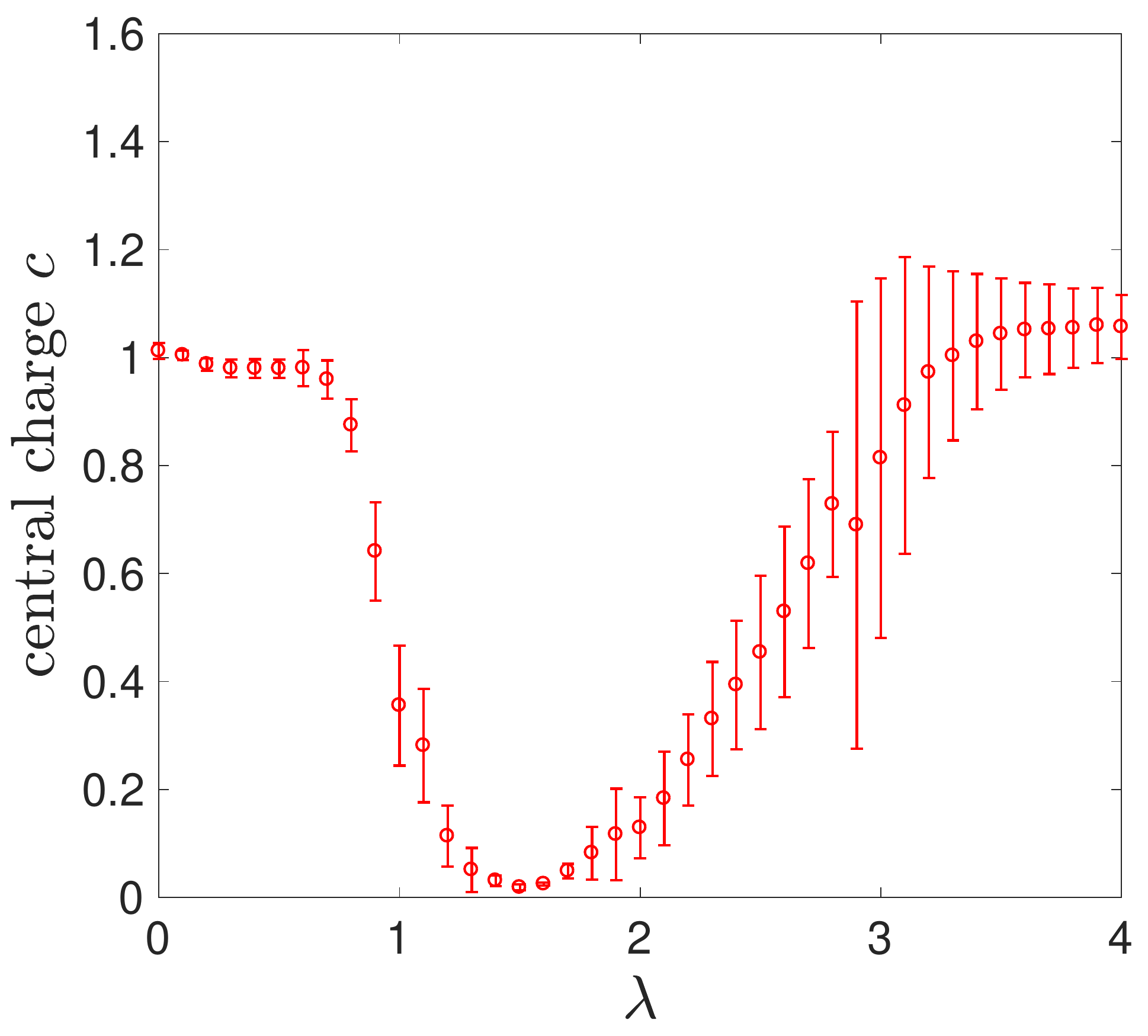}}
\caption{(a) von Neumann entropy versus the logarithm of the maximal correlation length.
Dash lines correspond to CFT predictions with $c=1$. 
(b) Central charge $c$ from linear fits of the data. 
Note, in the second critical phase, data with $\chi\ge 20D^2$ only are used in the fit.}
\label{fig:CFT_data}
\end{figure}

{\it Boundary CFT for $U(1)$ SL}.
The existence of the two critical SL phases (beyond the $\lambda=0$ and $\lambda=\infty$ points). 
is further supported by the analysis of the boundary state (see supplementary materials for details).
We find that the corresponding von Neumann entanglement entropy scales with the maximal correlation length $\xi_{\mathrm{B}}$ of the boundary state, 
when increasing $\chi$, like ${\bf{S}}_{vN}(\chi)\sim \frac{c}{6}\ln{\xi_{\mathrm{B}}(\chi)}$, as expected in
a (1+1) dimensional conformal field theory (CFT) with universal central charge $c$~\cite{Pollmann2009,Kjall2013}.
From the fits of Fig.~\ref{fig:CFT_data},  one gets $c=1.01(2)$ and $c=1.05(6)$,
consistent with a simple $c=1$ CFT. 

{\it Conclusion and outlook.}
Using a simple PEPS ansatz of a generalized RVB spin liquid, we have shown that (i) spin-$1/2$ topological SL 
with $C_{4v}$ point group symmetry and $SU(2)$ spin rotation symmetry exists on the square lattice and
(ii) criticality and nonbipartiteness are compatible. The topological phase observed here is naturally
connected to the $\mathbb{Z}_2$ gauge symmetry of the local tensor typical of an Ising gauge theory~\cite{Moessner2002,Fradkin2013}.
These properties are reminiscent of a classical interacting dimer model which interpolates between the square lattice and the triangular lattice by tuning a chemical potential in the diagonal bonds~\cite{Trousselet2007}.  In this case, 
by varying the temperature, a similar transition is seen separating a 
high-temperature short-range disordered dimer liquid to a critical KT low-temperature dimer phase.  In fact, the NN RVB phase can be mapped to
a classical {\it interacting} dimer model on the square lattice at finite temperature~\cite{Alet2005,Alet2006,Damle2012}. 
Although it is not clear how such a mapping could be extended once $\lambda\ne 0$, 
the similarity between the two phase diagrams suggests that both can be captured by the same long-wavelength height field theory~\cite{Henley1997,Alet2006,Trousselet2007},
giving rise to (continuous) KT phase transitions. Such a scenario is supported by the finding of a universal boundary
central charge $c=1$ in the two critical regions.

Finally, we note that the existence of a
SU(2)-invariant {\it local} parent Hamiltonian follows from the $\mathbb{Z}_2$-injectivity of the PEPS~\cite{Schuch2012}.
The latter at $\lambda\ne 0$ would be a ``deformation'' of the
parent Hamiltonian derived at $\lambda=0$ (see Supplementary Materials and Ref.~\cite{Fendley,Mambrini2015}), including e.g. (physically relevant) plaquette cyclic terms or other sorts of multi-spin interactions involving up to 6 sites. 

\begin{acknowledgements}
This project is supported by the TNSTRONG
ANR grant (French Research Council).  This work was granted access to the HPC resources of CALMIP 
supercomputing center under the allocation 2017-P1231. We acknowledge inspiring conversations with Sylvain Capponi, Pierre Pujol and Frank Pollmann.
JYC thanks Qing-Rui Wang, Jia-Wei Mei, Huan He and Hong-Hao Tu for many helpful discussions.

\end{acknowledgements}

\nocite{apsrev41Control}
\bibliographystyle{apsrev4-1}
\bibliography{squareZ2SL}

\end{document}

\title{Supplementary Materials for  ``Topological $\mathbb{Z}_2$ Resonating-Valence-Bond Spin Liquid on the Square Lattice''}

\author{Ji-Yao Chen}
\affiliation{Laboratoire de Physique Th\'eorique, C.N.R.S. and Universit\'e de Toulouse, 31062 Toulouse, France}
\author{Didier Poilblanc}
\affiliation{Laboratoire de Physique Th\'eorique, C.N.R.S. and Universit\'e de Toulouse, 31062 Toulouse, France}

\date{\today}
\maketitle

\section{Nonzero elements of tensor $\mathcal{A}_1^{(1,2)}$}
We list below the nonzero elements of tensor $\mathcal{A}_1^{(1,2)}$, reproduced from Ref.~\onlinecite{Mambrini2016}.
We use $\{ |0\rangle, |1\rangle, |2\rangle \}$ to label the $D=3$ virtual space basis with $\{ |0\rangle, |1\rangle \}$ for the virtual spin-$\frac{1}{2}$, $|2\rangle$ for the virtual spin-$0$, and use $\{ |0\rangle, |1\rangle \}$ for the physical spin-$\frac{1}{2}$. %The tensor indices are arranged as \emph{up, left, down, right} and \emph{physical} index. 
The tensor indices are arranged in $(u, l, d, r, s)$ order, see Fig.~\ref{fig:CTMRG}(a).

\begin{table}[htb]
\renewcommand\arraystretch{1.5}
\caption{The $\mathcal{A}_1^{(1)}$ NN RVB tensor.}
\label{table:A1_1}
\begin{tabularx}{0.48\textwidth}{l l}
	\hline
	\hline
	${\bf S}_z = +1/2$ \qquad\qquad\qquad & \qquad\qquad\qquad ${\bf S}_z = -1/2$ \\
	\hline
	$\mathcal{A}_1^{(1)}(0,2,2,2,0) = \frac{1}{2}$ \qquad\qquad & \qquad\qquad $\mathcal{A}_1^{(1)}(1,2,2,2,1) = \frac{1}{2}$ \\
	$\mathcal{A}_1^{(1)}(2,0,2,2,0) = \frac{1}{2}$ \qquad\qquad & \qquad\qquad $\mathcal{A}_1^{(1)}(2,1,2,2,1) = \frac{1}{2}$ \\
	$\mathcal{A}_1^{(1)}(2,2,0,2,0) = \frac{1}{2}$ \qquad\qquad & \qquad\qquad $\mathcal{A}_1^{(1)}(2,2,1,2,1) = \frac{1}{2}$ \\
	$\mathcal{A}_1^{(1)}(2,2,2,0,0) = \frac{1}{2}$ \qquad\qquad & \qquad\qquad $\mathcal{A}_1^{(1)}(2,2,2,1,1) = \frac{1}{2}$ \vspace{1mm}\\
	\hline
	\hline
\end{tabularx}
\end{table}

\begin{table}[hb]
\renewcommand\arraystretch{1.5}
\caption{The $\mathcal{A}_1^{(2)}$ long-range RVB tensor.}
\label{table:A1_2}
\begin{tabularx}{0.48\textwidth}{l l}
	\hline
	\hline
	${\bf S}_z = +1/2$ \qquad\qquad & \qquad\qquad ${\bf S}_z = -1/2$ \\
	\hline
	$\mathcal{A}_1^{(2)}(0,0,1,2,0) = -\frac{1}{2\sqrt{6}}$ \qquad\quad & \qquad $\mathcal{A}_1^{(2)}(0,1,1,2,1) = \frac{1}{2\sqrt{6}}$ \\
	$\mathcal{A}_1^{(2)}(0,0,2,1,0) = -\frac{1}{2\sqrt{6}}$ \qquad\quad & \qquad $\mathcal{A}_1^{(2)}(0,1,2,1,1) = -\frac{1}{\sqrt{6}}$ \\
	$\mathcal{A}_1^{(2)}(0,1,0,2,0) = \frac{1}{\sqrt{6}}$ \qquad\quad & \qquad $\mathcal{A}_1^{(2)}(0,2,1,1,1) = \frac{1}{2\sqrt{6}}$ \\
	$\mathcal{A}_1^{(2)}(0,1,2,0,0) = -\frac{1}{2\sqrt{6}}$ \qquad\quad & \qquad $\mathcal{A}_1^{(2)}(1,0,1,2,1) = -\frac{1}{\sqrt{6}}$ \\
	$\mathcal{A}_1^{(2)}(0,2,0,1,0) = \frac{1}{\sqrt{6}}$ \qquad\quad & \qquad $\mathcal{A}_1^{(2)}(1,0,2,1,1) = \frac{1}{2\sqrt{6}}$ \\
	$\mathcal{A}_1^{(2)}(0,2,1,0,0) = -\frac{1}{2\sqrt{6}}$ \qquad\quad & \qquad $\mathcal{A}_1^{(2)}(1,1,0,2,1) = \frac{1}{2\sqrt{6}}$ \\
	$\mathcal{A}_1^{(2)}(1,0,0,2,0) = -\frac{1}{2\sqrt{6}}$ \qquad\quad & \qquad $\mathcal{A}_1^{(2)}(1,1,2,0,1) = \frac{1}{2\sqrt{6}}$ \\
	$\mathcal{A}_1^{(2)}(1,0,2,0,0) = \frac{1}{\sqrt{6}}$ \qquad\quad & \qquad $\mathcal{A}_1^{(2)}(1,2,0,1,1) = \frac{1}{2\sqrt{6}}$ \\
	$\mathcal{A}_1^{(2)}(1,2,0,0,0) = -\frac{1}{2\sqrt{6}}$ \qquad\quad & \qquad $\mathcal{A}_1^{(2)}(1,2,1,0,1) = -\frac{1}{\sqrt{6}}$ \\
	$\mathcal{A}_1^{(2)}(2,0,0,1,0) = -\frac{1}{2\sqrt{6}}$ \qquad\quad & \qquad $\mathcal{A}_1^{(2)}(2,0,1,1,1) = \frac{1}{2\sqrt{6}}$ \\
	$\mathcal{A}_1^{(2)}(2,0,1,0,0) = \frac{1}{\sqrt{6}}$ \qquad\quad & \qquad $\mathcal{A}_1^{(2)}(2,1,0,1,1) = -\frac{1}{\sqrt{6}}$ \\
	$\mathcal{A}_1^{(2)}(2,1,0,0,0) = -\frac{1}{2\sqrt{6}}$ \qquad\quad & \qquad $\mathcal{A}_1^{(2)}(2,1,1,0,1) = \frac{1}{2\sqrt{6}}$ \vspace{1mm}\\	
	\hline
	\hline
\end{tabularx}
\end{table}

The $SU(2)$ symmetry and $C_{4v}$ lattice symmetry obeyed by the tensor $\mathcal{A}_1^{(1,2)}$ can be directly checked using the tensor coefficients shown in Tables \ref{table:A1_1} and~\ref{table:A1_2}.

The long-range RVB state can be seen as a fluctuating (bond-doped) Affeck-Kennedy-Lieb-Tasaki (AKLT) state where virtual (spin-1/2 or spin-0) degrees of freedom attached to the sites pair up in either singlet or ``empty'' bonds. As in the AKLT picture the virtual spins are projected onto the physical spin degrees of freedom (spin 1/2 here, instead of spin-2 for the two-dimensional AKLT state).

\section{CTMRG and correlation function}
\begin{figure}[htb]
\centering
	\subfloat[]{\includegraphics[width=20mm, height = 20mm]{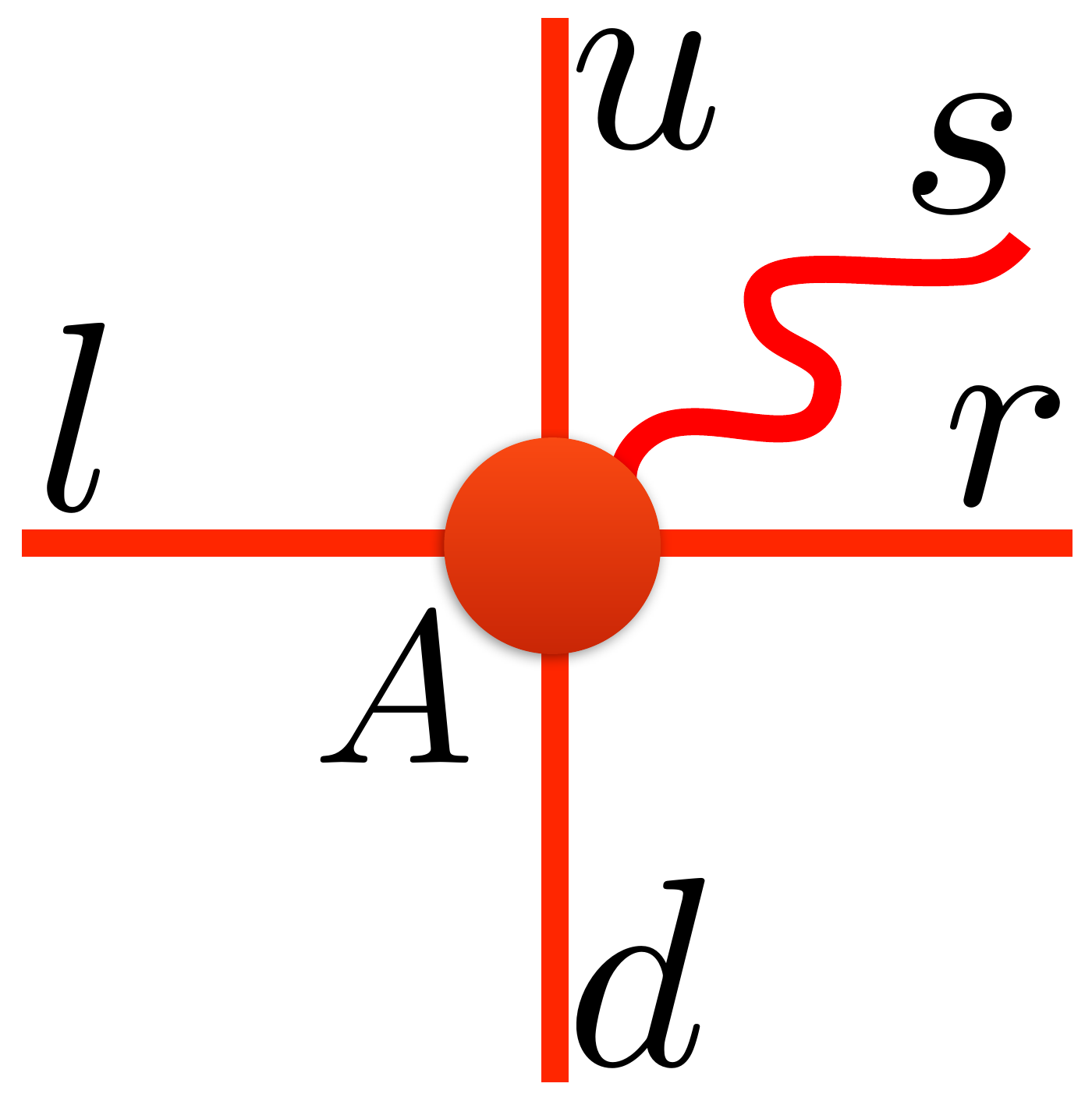}\label{fig:tensorA1}}\hspace{10mm}
	\subfloat[]{\includegraphics[width=30mm, height = 30mm]{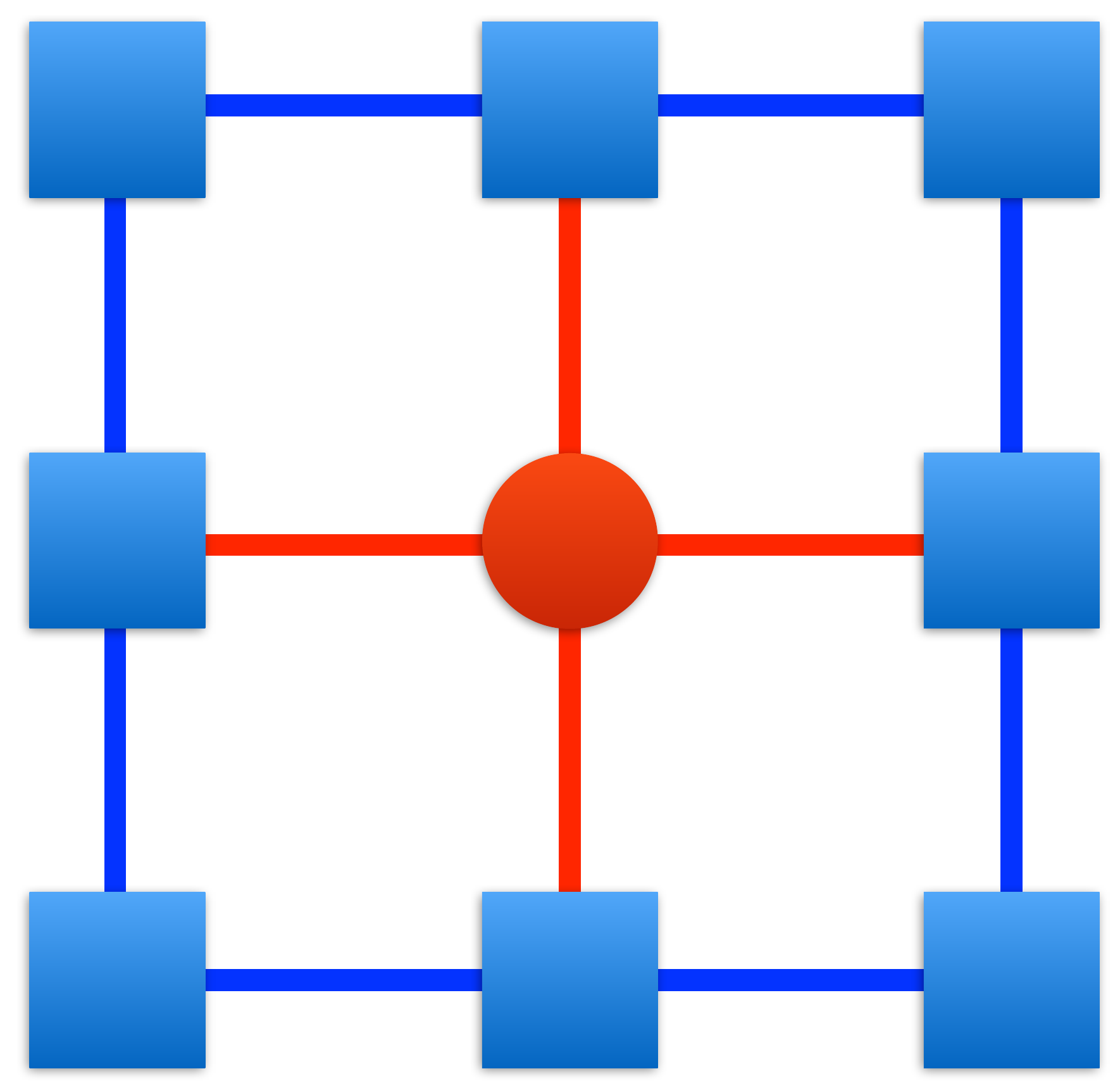}\label{fig:environment}}\\
	\subfloat[]{\includegraphics[width=18mm, height = 18mm]{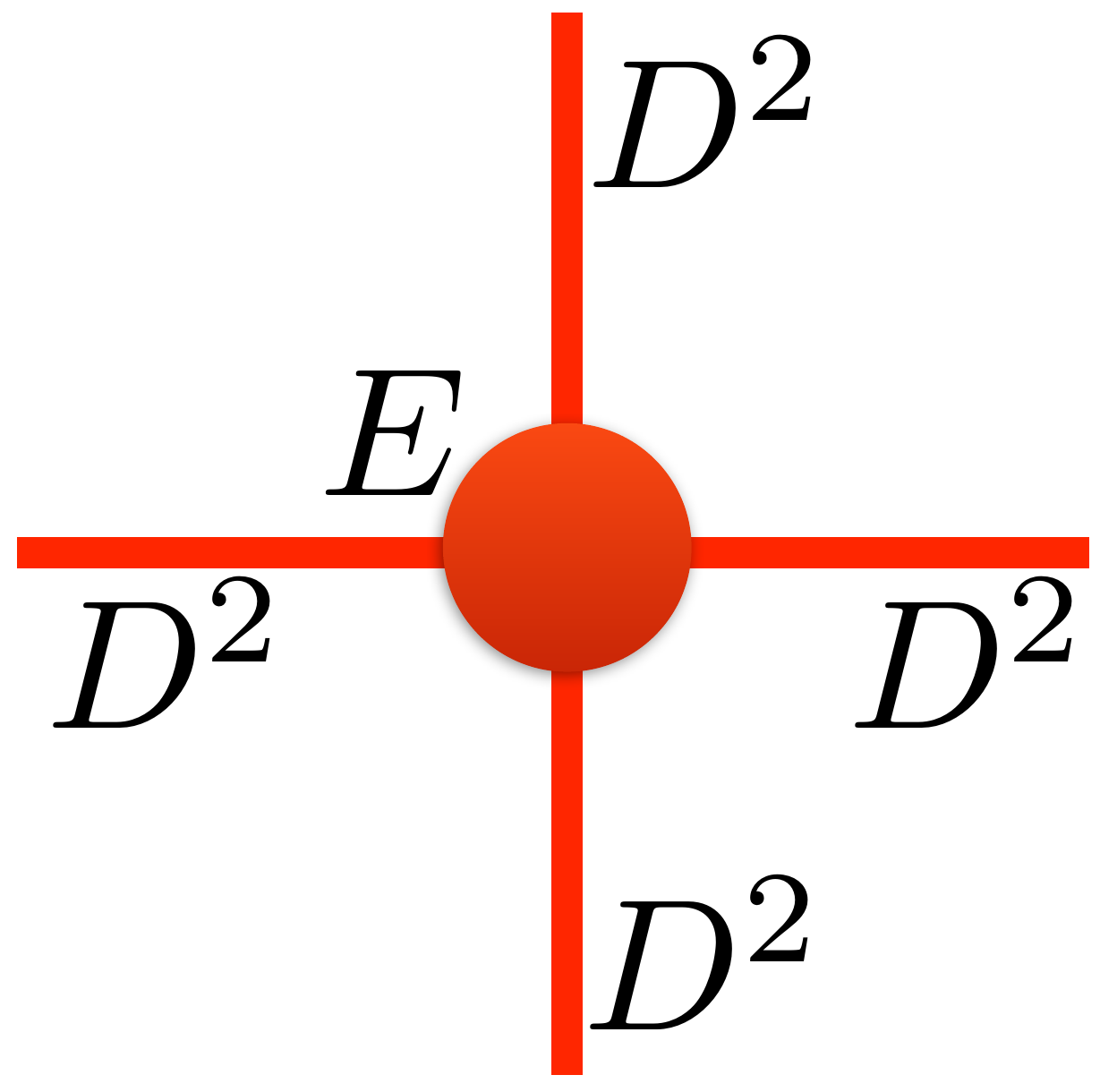}\label{fig:doubleTensorE}}\hfill
	\subfloat[]{\includegraphics[width=17mm, height = 17mm]{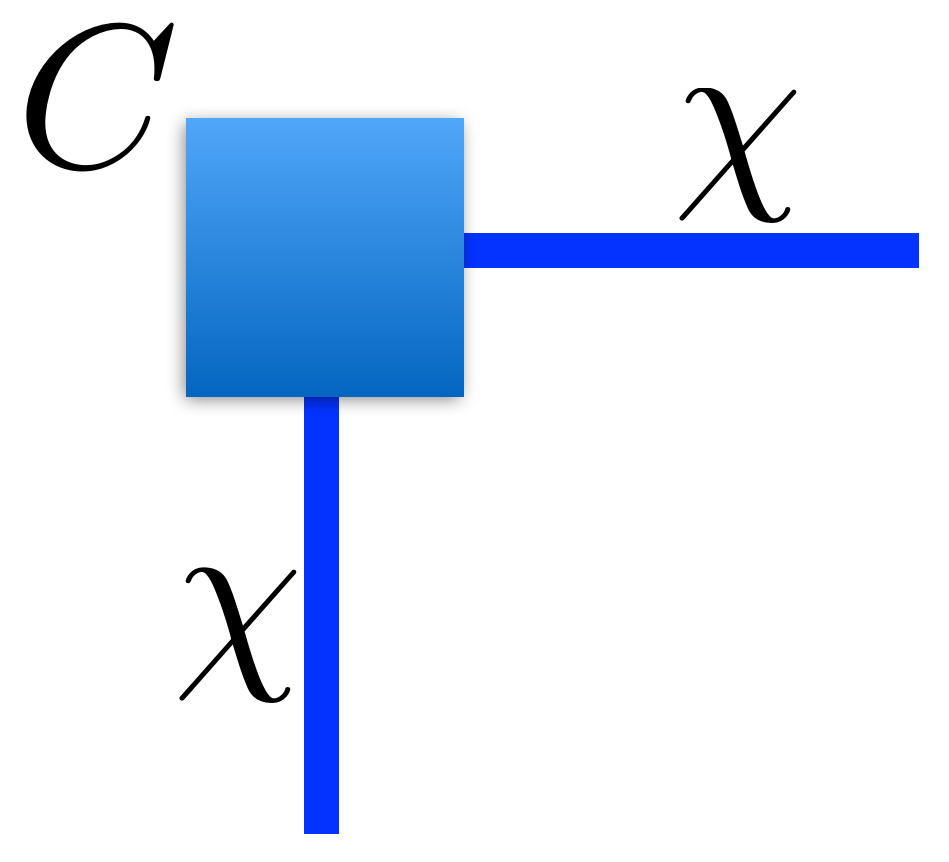}\label{fig:cornMatC}}\hfill
	\subfloat[]{\includegraphics[width=20mm, height = 16mm]{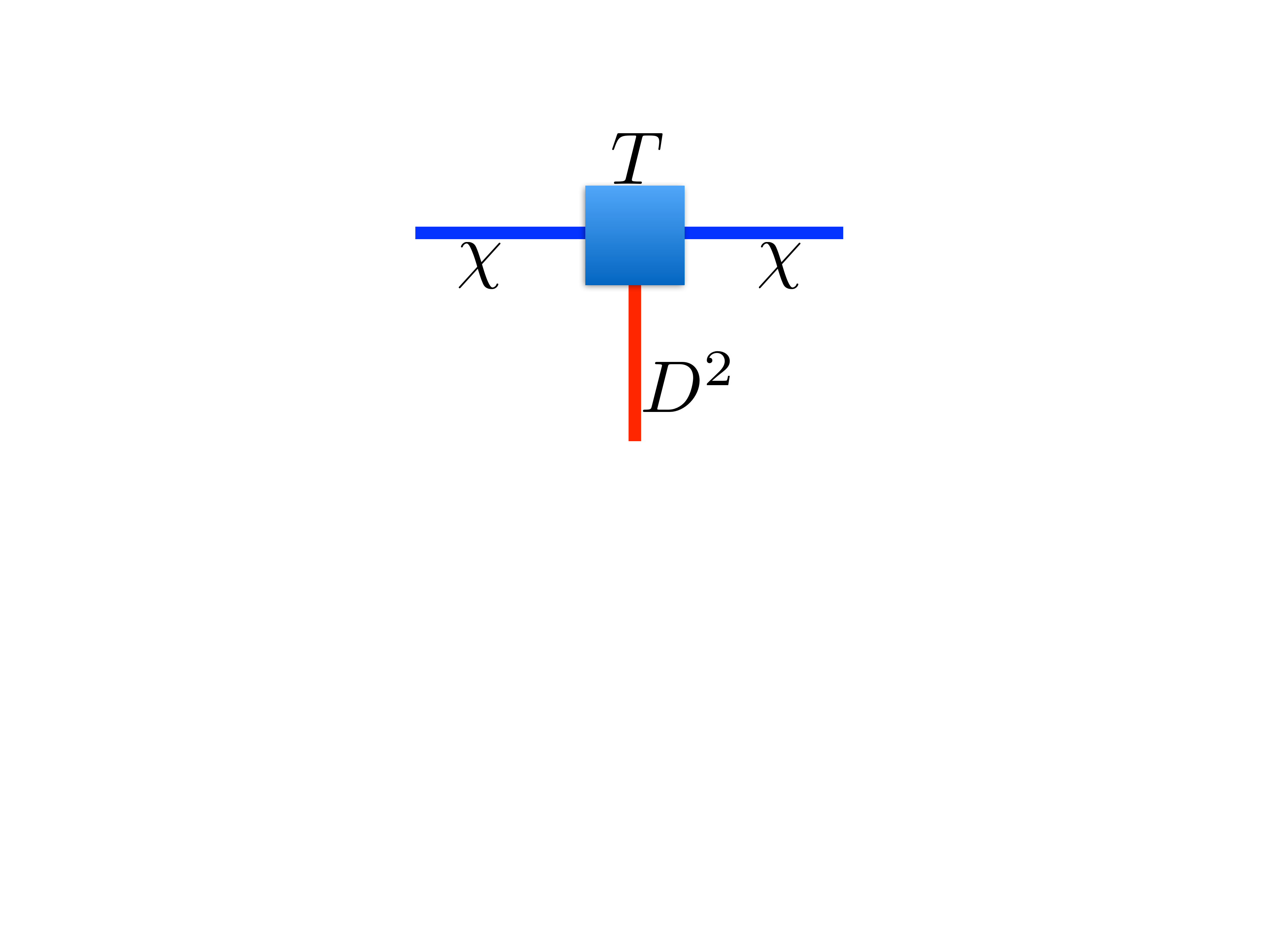}\label{fig:tranMatT}}
\caption{One-site unit cell and its environment tensors in CTMRG. 
In (b), red filled circle represents the double tensor $E$, which is obtained by contracting the physical index of PEPS tensor $A$: $E = \mathrm{tr}(A^{s*}A^s)$. 
The environment is constructed by corner matrix $C$ and transfer matrix $T$, with environment bond dimension $\chi$.}
\label{fig:CTMRG}
\end{figure}

To calculate physical observables, we apply the corner transfer matrix renormalization group (CTMRG) method. Note that the PEPS we are dealing with is constructed by a single tensor which has $C_{4v}$ lattice symmetry. Therefore we can use the simplified CTMRG method introduced in Ref.~\onlinecite{Poilblanc2017}, and approximate 
 the environment of the one-site unit cell with diagonal corner matrix $C$ and transfer matrix $T$, whose size is $\chi\times\chi$ and $\chi\times D^2\times\chi$. Here $\chi=kD^2 (k\in\mathbb{N}^+)$ is the environment bond dimension, which controls the precision of this method. 
See Fig.~\ref{fig:CTMRG} for graphic representation.

\begin{figure}[hbt]
\centering
	\subfloat[]{\includegraphics[width=38mm, height = 21mm]{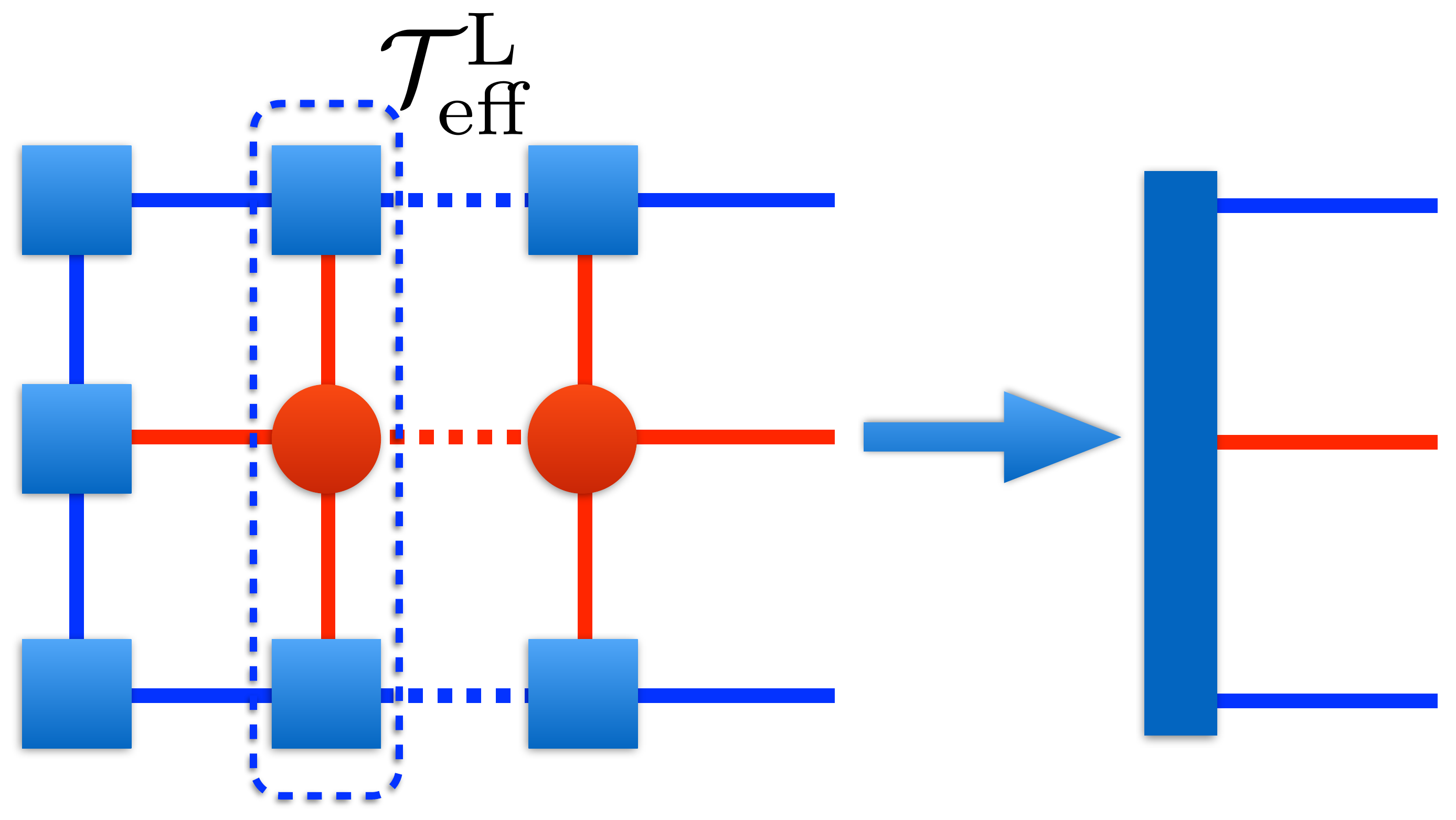}\label{fig:eigenVectorL}}\hfill
	\subfloat[]{\includegraphics[width=38mm, height = 28mm]{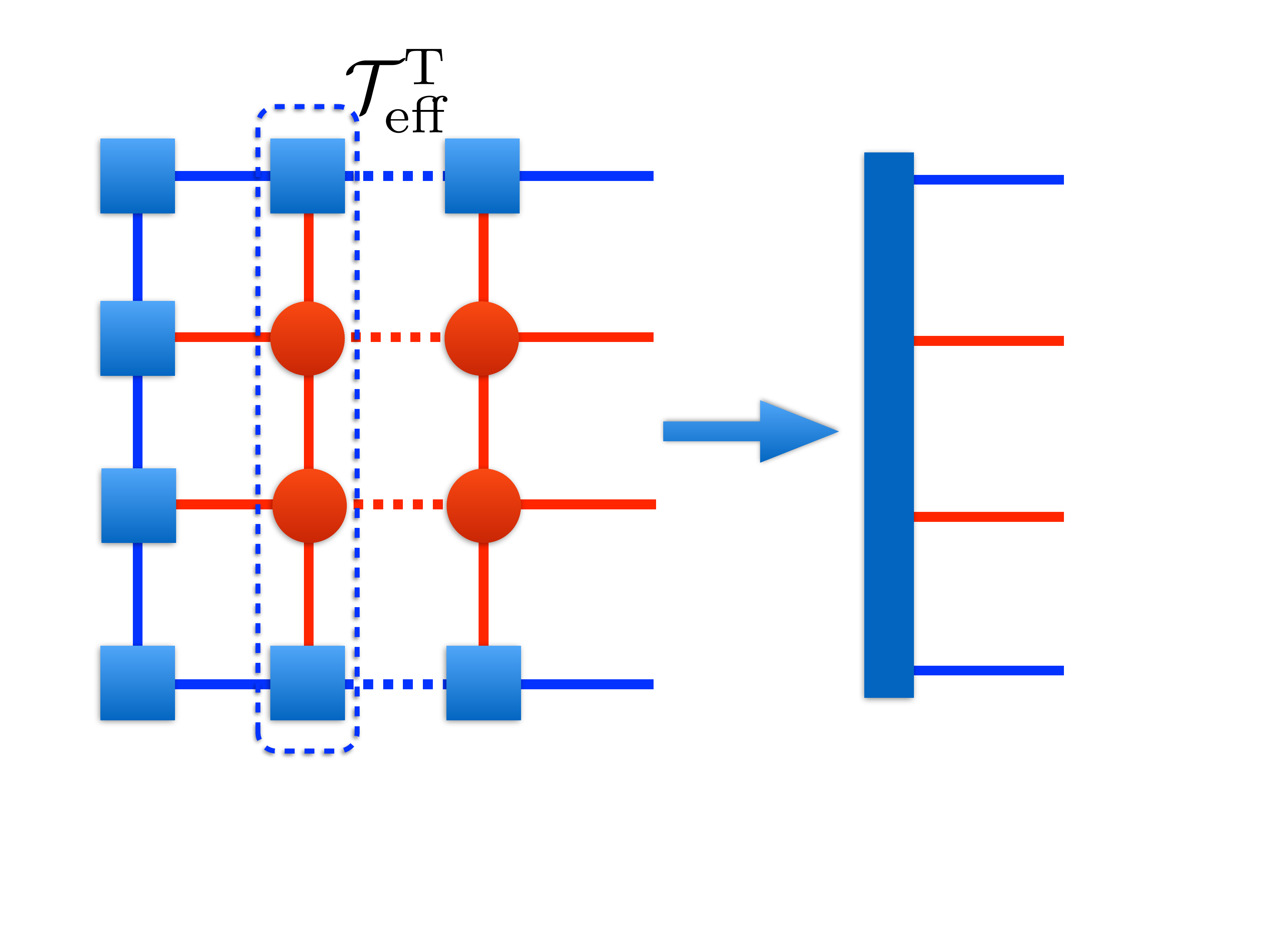}\label{fig:eigenVectorT}}\\
	\subfloat[]{\includegraphics[width=39mm, height = 23mm]{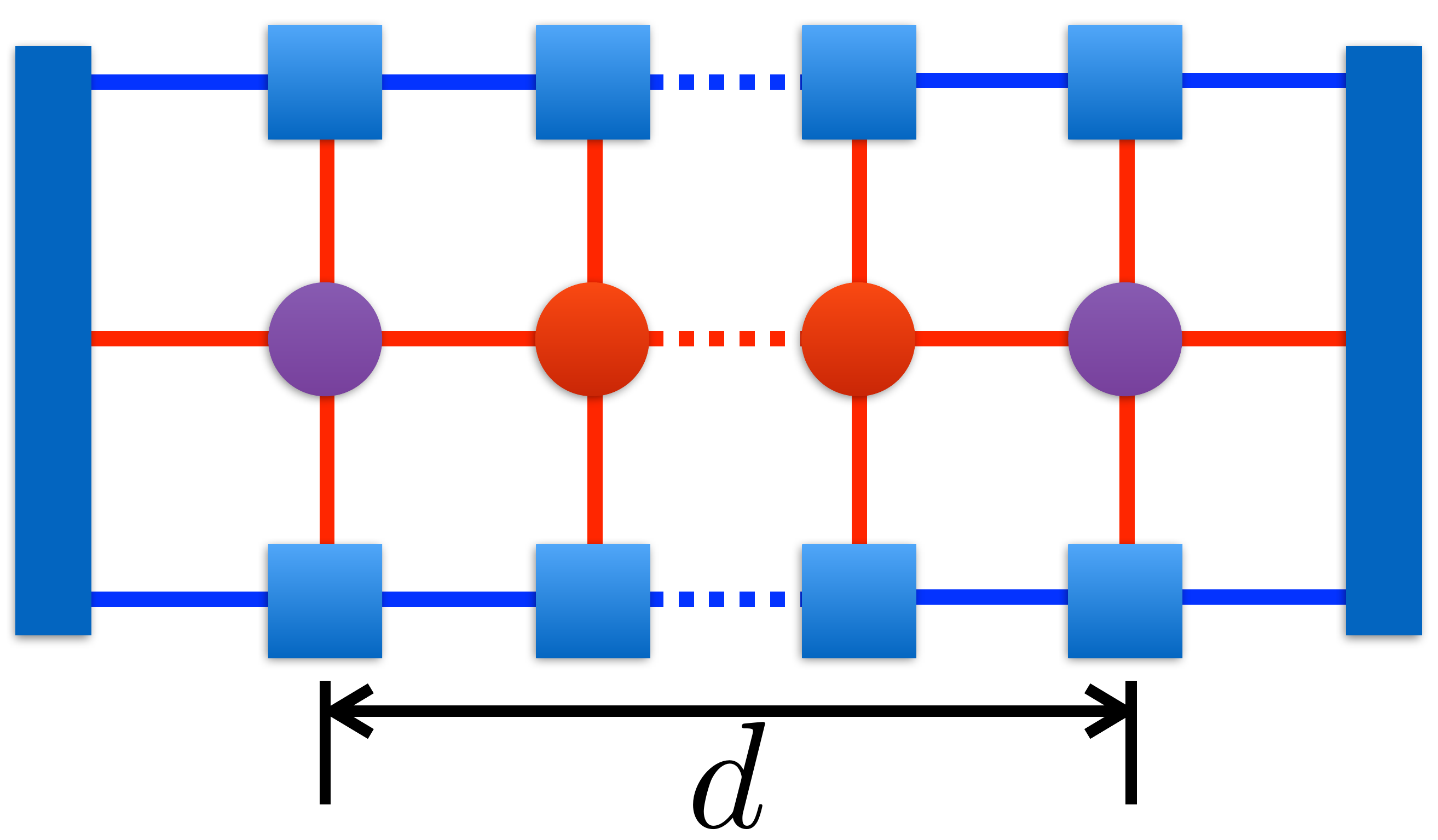}\label{fig:ssCorrFunc}}\hfill
	\subfloat[]{\includegraphics[width=45mm, height = 23mm]{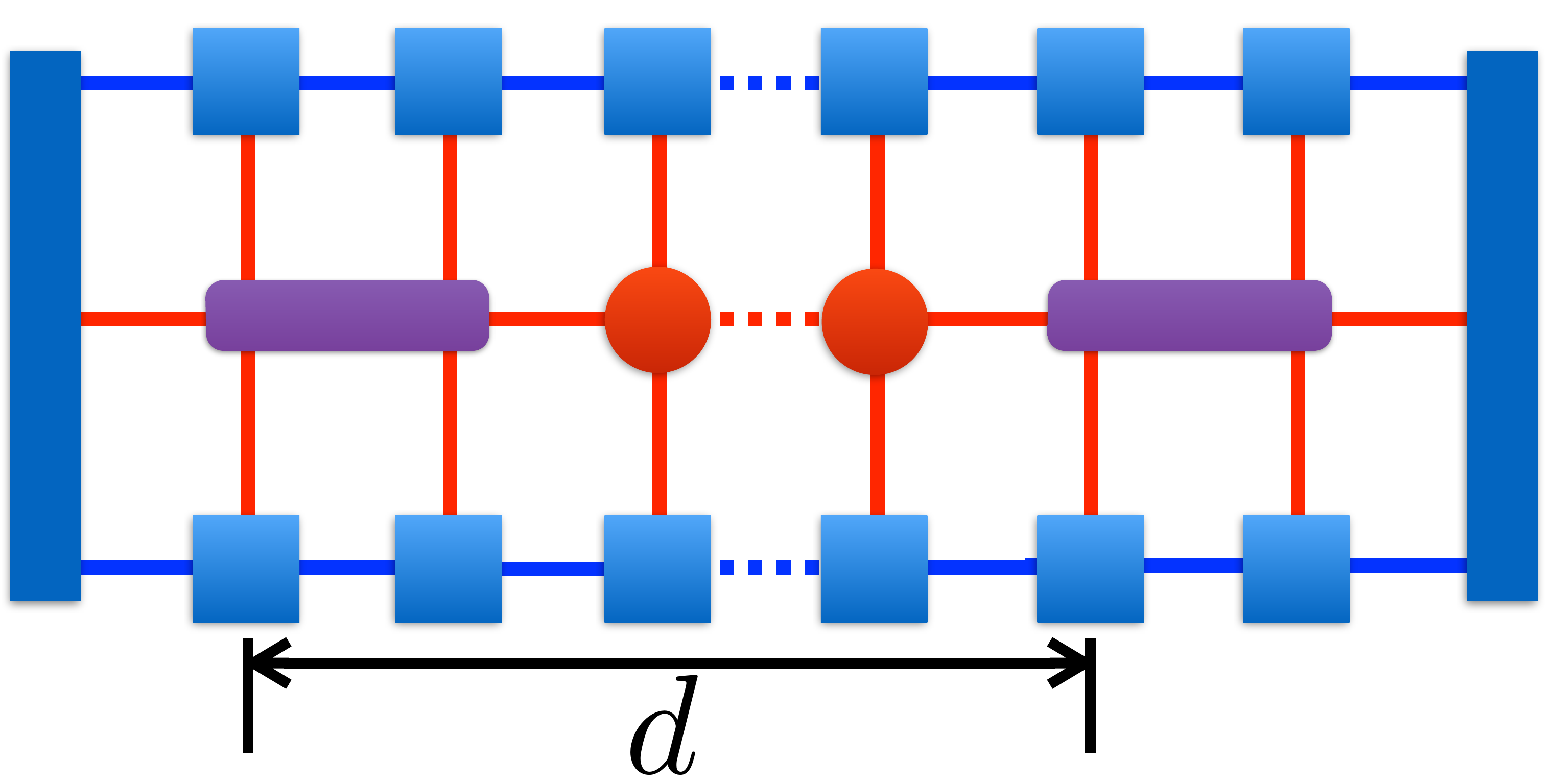}\label{fig:lddCorrFunc}}\\
	\subfloat[]{\includegraphics[width=45mm, height = 33mm]{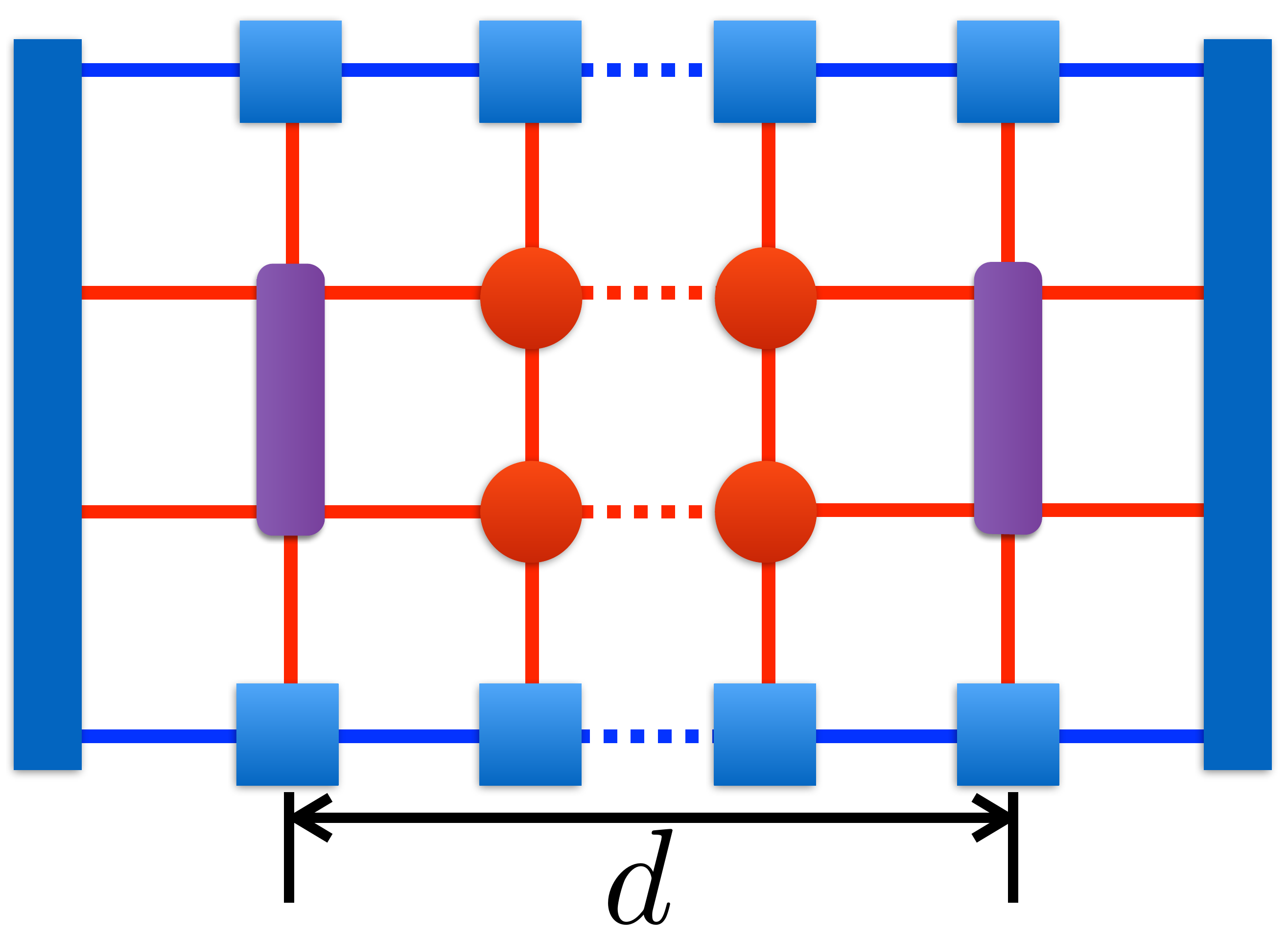}\label{fig:tddCorrFunc}}
\caption{Network for correlation function. First we use power iteration to calculate leading eigenvector of effective transfer matrix $\mathcal{T}_{\mathrm{eff}}^{\mathrm{L,T}}$, as shown in (a) and (b), which serve as boundary vector in (c), (d), (e). Then various correlation functions can be obtained by putting corresponding physical operators on suitable sites, represented by purple symbols.}
\label{fig:correlator}
\end{figure}

With the environment tensors $C$ and $T$, it is straightforward to calculate various correlation functions, \emph{e.g.}, spin-spin, longitudinal and transverse dimer-dimer correlations.
The networks to be contracted is similar to those in the matrix product state calculation.
When the correlation function decays very slowly, as the dimer-dimer correlations for the NN RVB state, it is important to obtain the correlation function in the extremely large distance with high precision, which will be vital to extract correlation length (under finite $\chi$) and critical exponent. For this, we first calculate the leading eigenvector of the corresponding transfer matrix, which can be achieved through a few power iteration steps, as shown in Fig.~\ref{fig:correlator}(a) and \ref{fig:correlator}(b). 
%Fig.~\ref{fig:correlator}(a) and (b).
And all other setups follows from Ref.~\onlinecite{Poilblanc2017}. The networks for spin-spin, longitudinal (transverse) dimer-dimer correlations are shown in Fig.~\ref{fig:correlator}(c) and \ref{fig:correlator}(d) (\ref{fig:correlator}(e)), where the corresponding leading eigenvector serves as boundary.

\begin{figure}[hbt]
\centering
	\includegraphics[width=69mm,height=70mm]{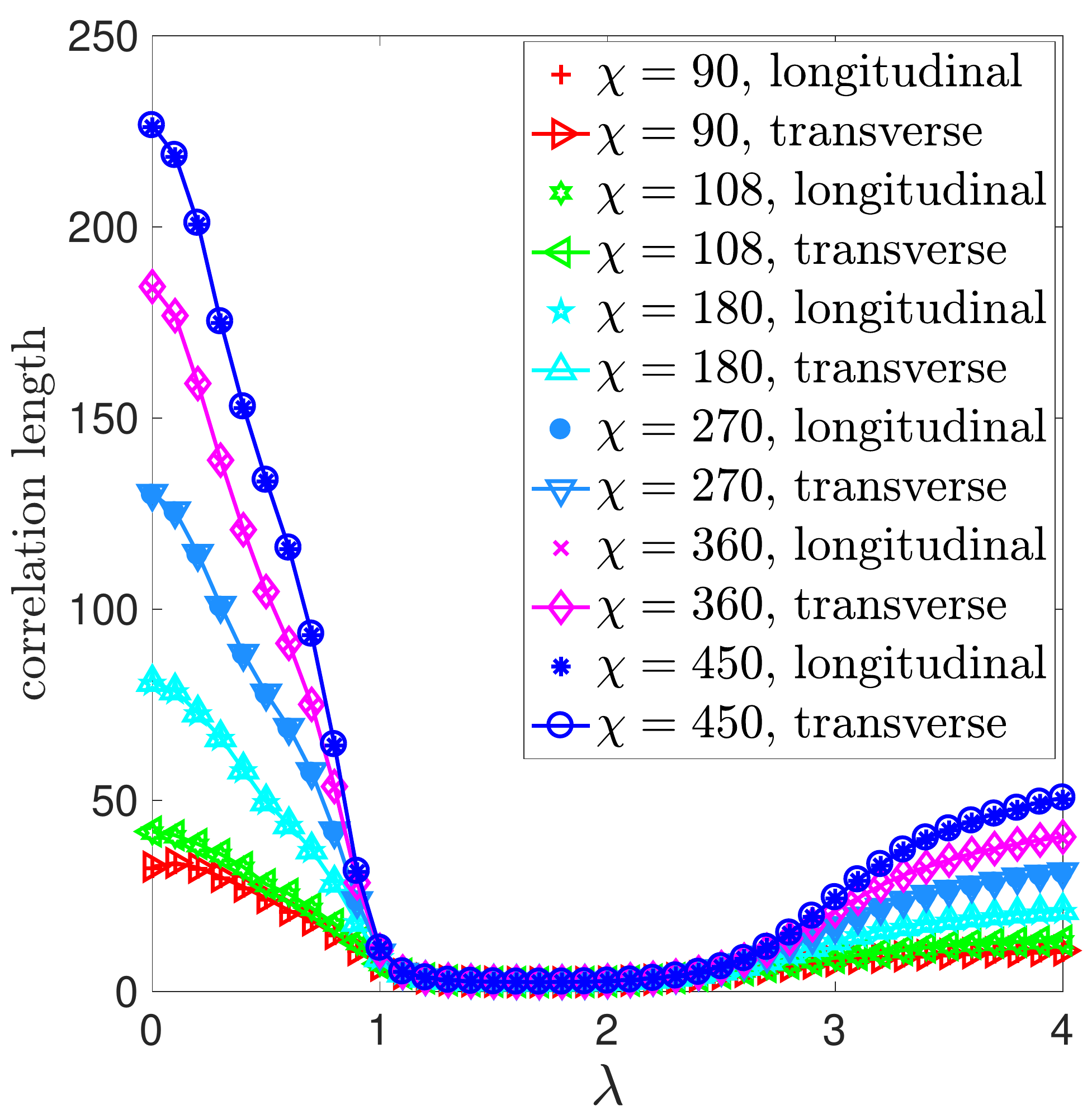}
\caption{$\xi^{\mathrm{L,T}}_{\mathrm{max}}$ versus $\lambda$, obtained from leading eigenvalues of effective transfer matrix. See Eq.~\eqref{eq:corrLengthTM}.}
\label{fig:corrLength_TM}
\end{figure}

One can determine the largest correlation length $\xi^{\mathrm{L,T}}_{\mathrm{max}}$ of the network from 
leading eigenvalues of effective transfer matrix $\mathcal{T}_{\mathrm{eff}}^{\mathrm{L,T}}$, which is hermitian due to reflection symmetry:
\begin{equation}\label{eq:corrLengthTM}
\xi^{\mathrm{L,T}}_{\mathrm{max}} = -\frac{1}{\mathrm{ln}|\lambda_2^{\mathrm{L,T}}/\lambda_1^{\mathrm{L,T}}|},
\end{equation}
where $\lambda_{1(2)}^{\mathrm{L,T}}$ is the first (second) largest eigenvalue of effective transfer matrix $\mathcal{T}_{\mathrm{eff}}^{\mathrm{L,T}}$.
As shown in Fig.~\ref{fig:corrLength_TM}, $\xi^{\mathrm{L,T}}_{\mathrm{max}}$ is close to the correlation length $\xi_{\mathrm{D}}^{\mathrm{L,T}}$, 
obtained from exponential fitting of longitudinal/transverse dimer-dimer correlation functions.
%except in the parameter region where dimer-dimer correlation length is too large to fit with a super good precision.
%This indicates that the spin singlet sector has the largest correlation length within all sectors.
Moreover, $\xi^{\mathrm{L,T}}_{\mathrm{max}}$ also shows divergence behavior with increasing $\chi$ in the two $U(1)$ phases 
and saturates to a finite value in the $\mathbb{Z}_2$ phase.

\section{Tensor renormalization group method and modular matrix}
To obtain the topological characteristics, we have implemented the tensor renormalization group (TRG) method to compute the modular matrices.
The PEPS we are working on has $\mathbb{Z}_2$ gauge symmetry except at $\lambda=0,\infty$ where $\mathbb{Z}_2$ is enlarged to $U(1)$. On the PEPS tensor level, the nontrivial gauge transformation is represented by $Z=\mathrm{diag}(-1, -1, 1)$, whose action on the four virtual indices of $\mathcal{A}$ only induces a minus sign.
Therefore we can use the method in Ref.~\onlinecite{He2014}, through acting different $\mathbb{Z}_2$ gauge transformation on the boundary of the double layer tensor network to get the modular matrices.

\begin{figure}[htb]
\centering
\raggedright
	\begin{minipage}[c]{0.285\textwidth}
	\subfloat[]{\includegraphics[width=45mm, height = 45mm]{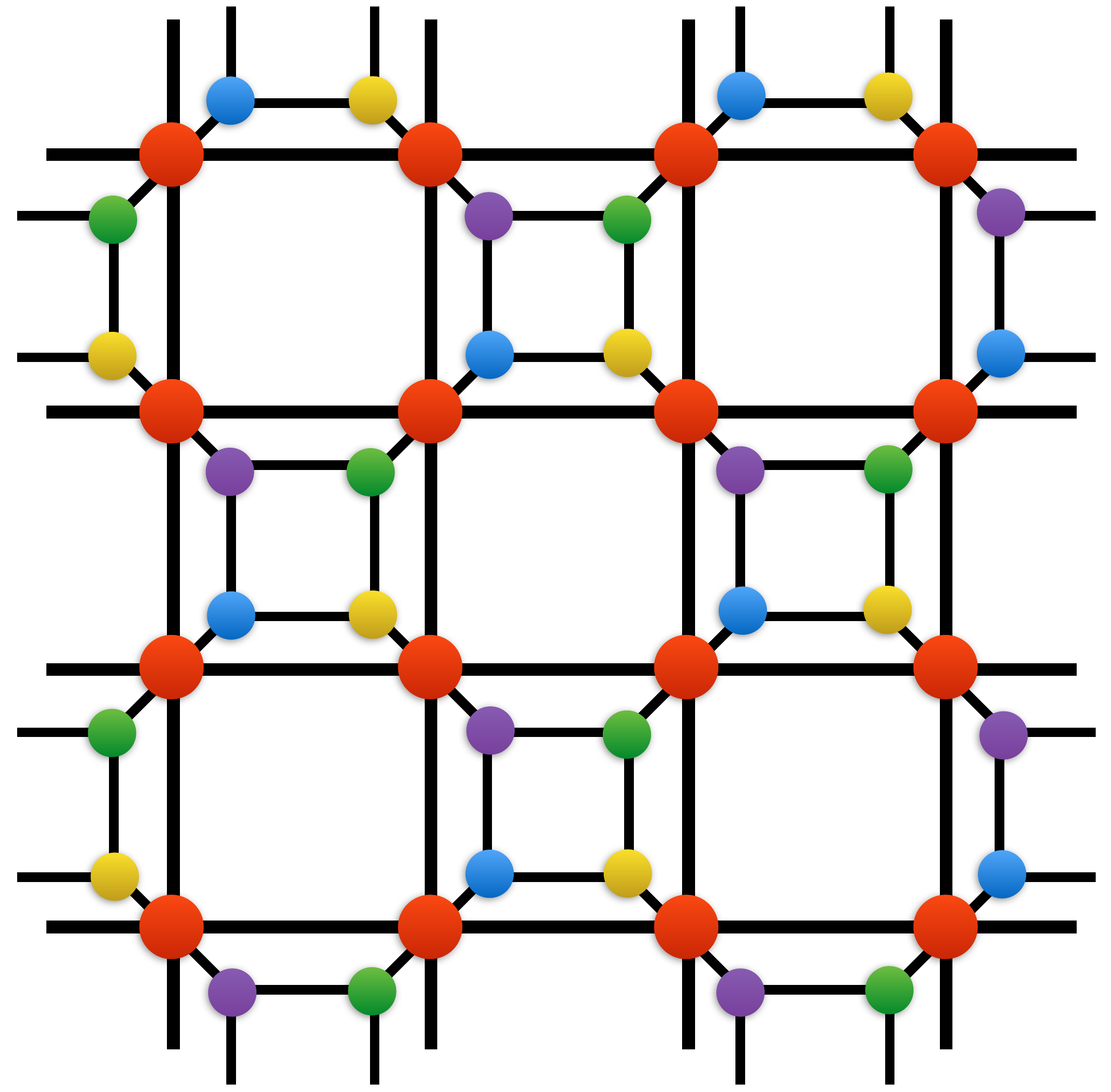}\label{fig:oneStepTRG}}
	\end{minipage}
	\begin{minipage}[c]{0.12\textwidth}
	\subfloat[]{\includegraphics[width=32mm, height = 32mm]{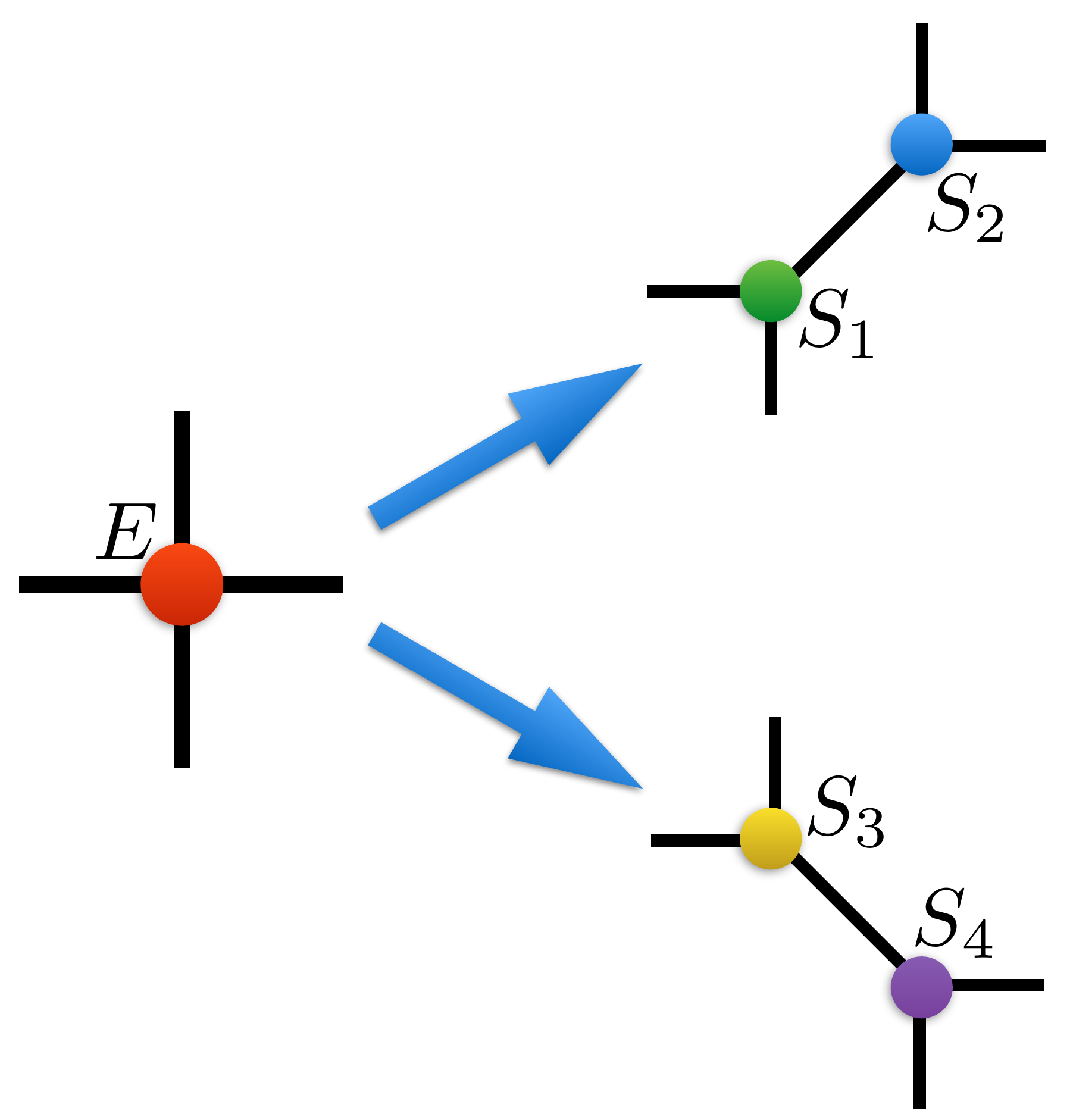}\label{fig:TRG_decompose}}\\
	\subfloat[]{\includegraphics[width=35mm, height = 15mm]{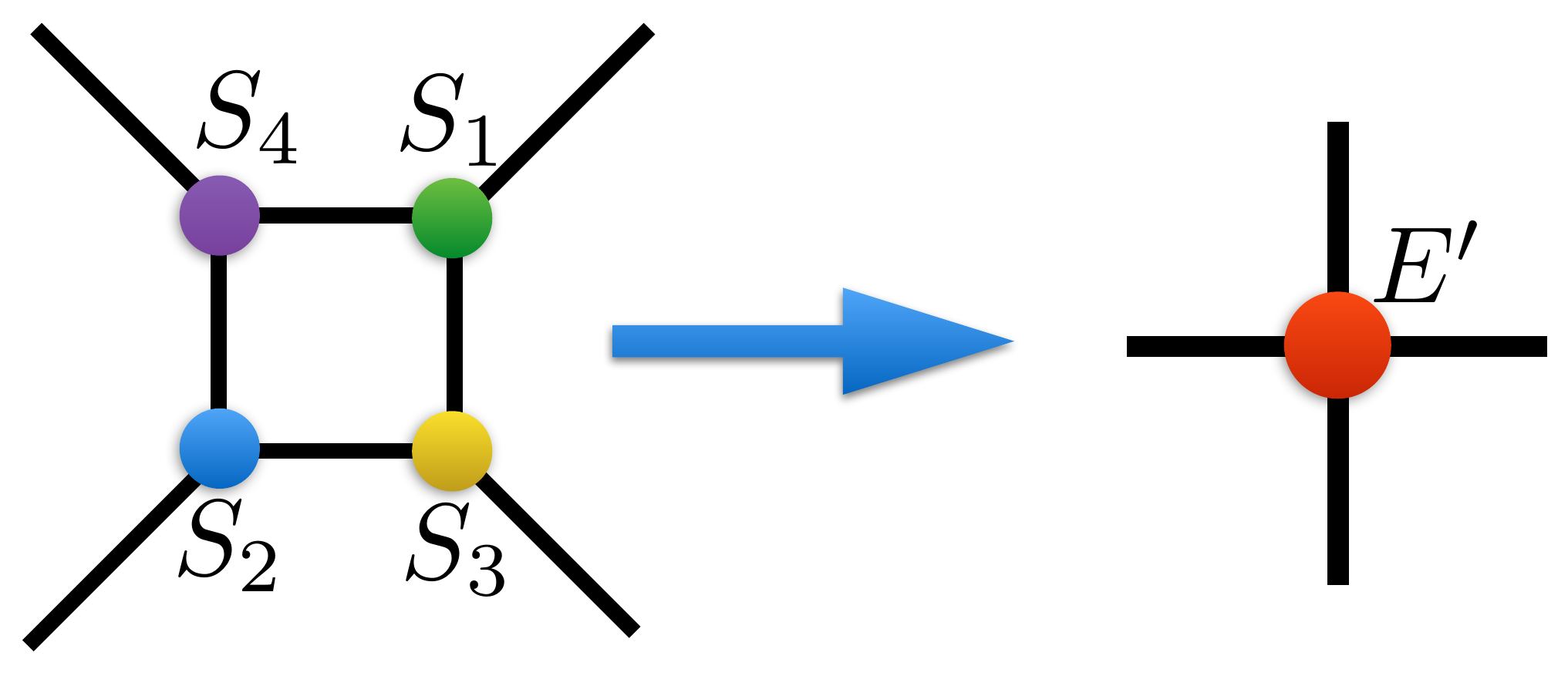}\label{fig:TRG_recombine}}
	\end{minipage}
\caption{Tensor renormalization group.
In (b), we first decompose the double tensor $E$ into two three-index tensors ${\bf S}_{1,2}$ or ${\bf S}_{3,4}$, depending on the sublattice.
Then we contract the four tensors ${\bf S}_{1,2,3,4}$ on the small square to get new double tensor $E'$, according to the order shown in (c).}
\label{fig:TRG}
\end{figure}

One typical TRG step on square lattice is shown in Fig.~\ref{fig:TRG}, where we first decompose the double tensor $E$ on every site into two three-index tensors ${\bf S}_{1,2}$ or ${\bf S}_{3,4}$ depending on the sublattice, using singular value decompostion (SVD).
Then we contract the four tensors ${\bf S}_{1,2,3,4}$ on the small square and obtain a new double tensor. 
Iterating in this way, we can obtain the fixed point double tensor. 

The controlling parameter of TRG is 
the bond dimension $\chi$ of the double tensor, which grows up in every iteration.
Therefore we have to truncate it to a finite value. 
Moreover, the $\mathbb{Z}_2$ gauge symmetry needs to be preserved during SVD, either through block SVD method~\cite{He2014} which we use here, or more advanced technique~\cite{Mei2017}.

\begin{figure}[htb]
\centering
	\subfloat[]{\includegraphics[width=30mm, height = 30mm]{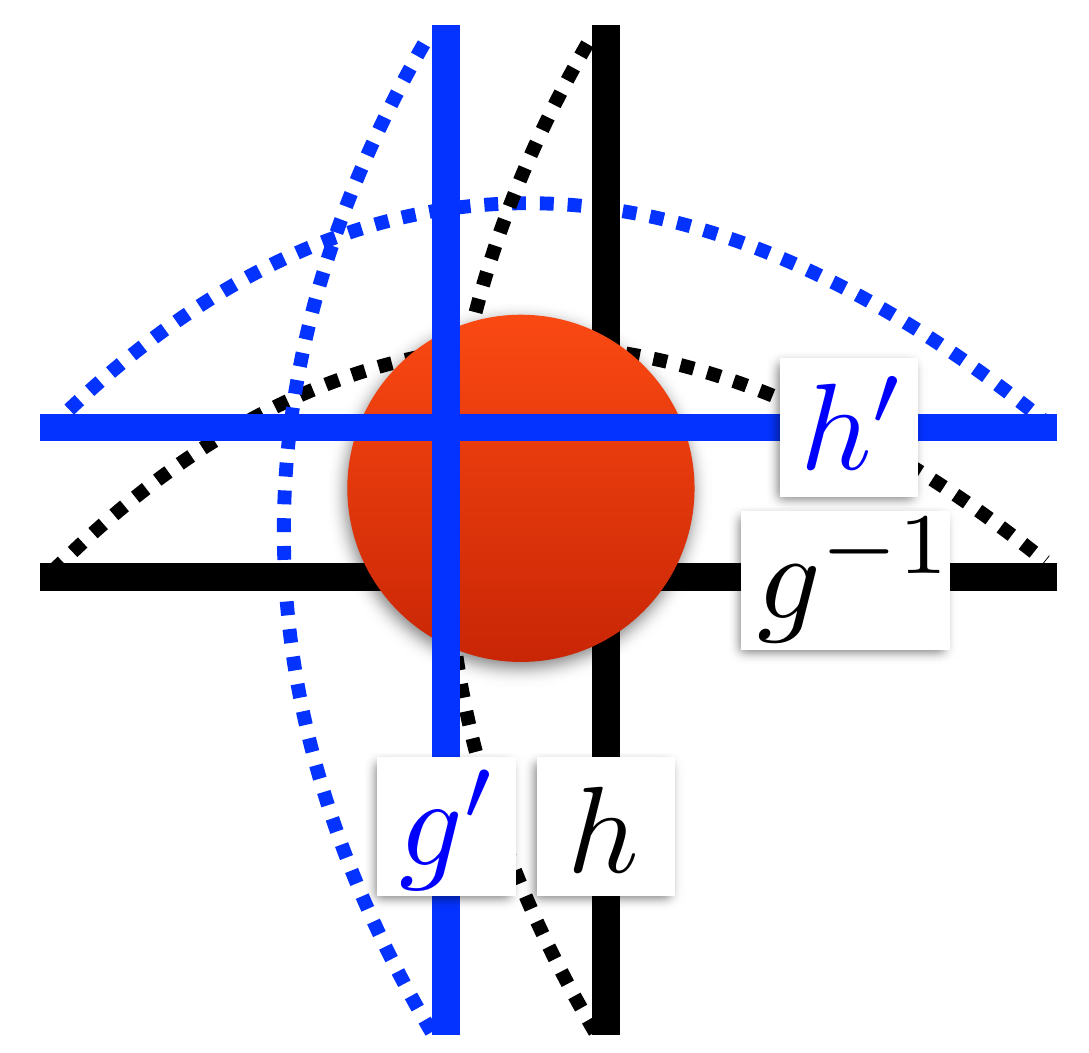}\label{fig:modularS}}\hfill
	\subfloat[]{\includegraphics[width=30mm, height = 30mm]{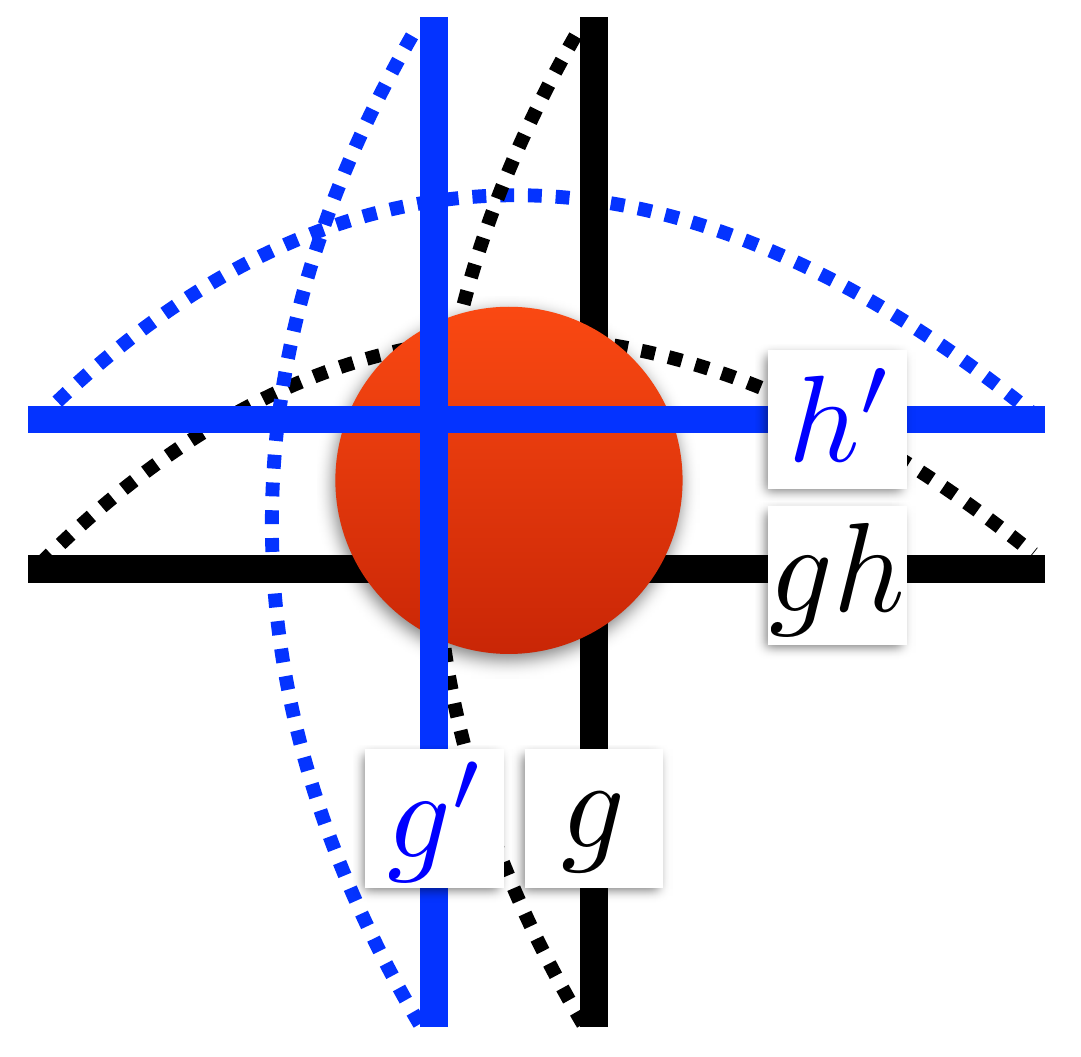}\label{fig:modularT}}\\
\caption{After one step TRG iteration, we put the double tensor on torus to compute modular matrices. 
(a) represents modular matrix $S$ element $S(g',h',g,h)$, where $g',h'$ acting on the top layer and $g^{-1},h$ acting on the bottom layer. 
And similarly with (b) for modular matrix $T$.}
\label{fig:modular_matrix}
\end{figure}

After every iteration, we put the double tensor on a torus, and adding various gauge transformations to compute the modular matrices. More explicitly, we have:
\begin{equation}
\begin{split}
&\langle \psi(g',h')|\hat{S}|\psi(g,h)\rangle = \langle\psi(g',h')|\psi(h,g^{-1})\rangle,\\
&\langle \psi(g',h')|\hat{T}|\psi(g,h)\rangle = \langle\psi(g',h')|\psi(g,gh)\rangle,
\end{split}
\end{equation}
which have also been shown graphically shown in Fig.~\ref{fig:modular_matrix}. Here $g',g',g,h$ are elements of the $\mathbb{Z}_2$ group.

\section{Boundary state of the PEPS}

The mapping between PEPS and classical statistical models is most easily seen by noting that the PEPS norm can be viewed as a partition function:
$\langle \psi(\mathcal{A})|\psi(\mathcal{A})\rangle = \mathcal{Z}_{\mathcal{A}}$. With the original physical indices contracted, the network associated to the partition function
is formed by the double tensor $E$, shown in Fig.~\ref{fig:CTMRG}(c).

\begin{figure}[htb]
\centering
\raggedright
	\begin{minipage}[l]{0.3\textwidth}
	\vspace{1mm}
	\subfloat[]{\includegraphics[width = 50mm, height = 12mm]{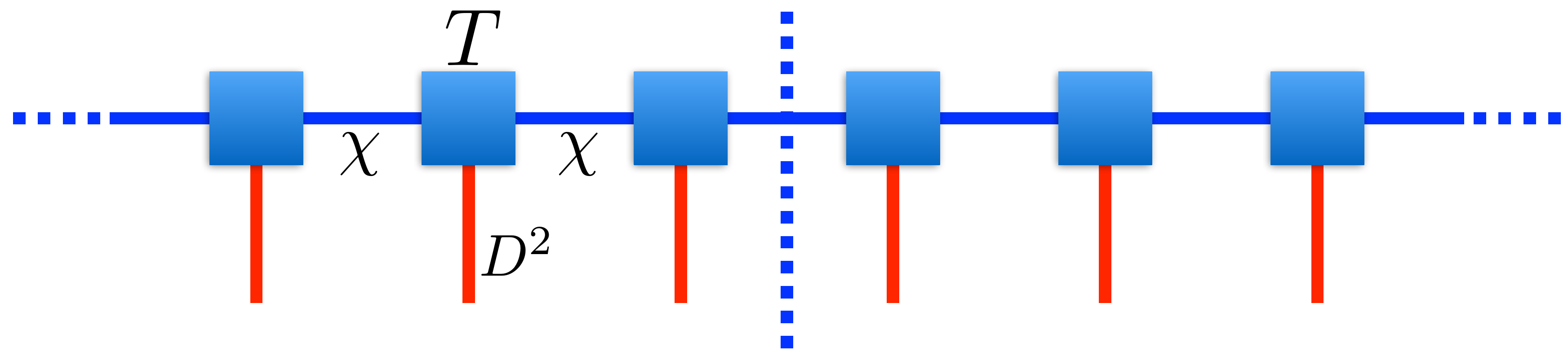}}\\
	\vspace{2mm}
	\subfloat[]{\includegraphics[width = 50mm, height = 11.5mm]{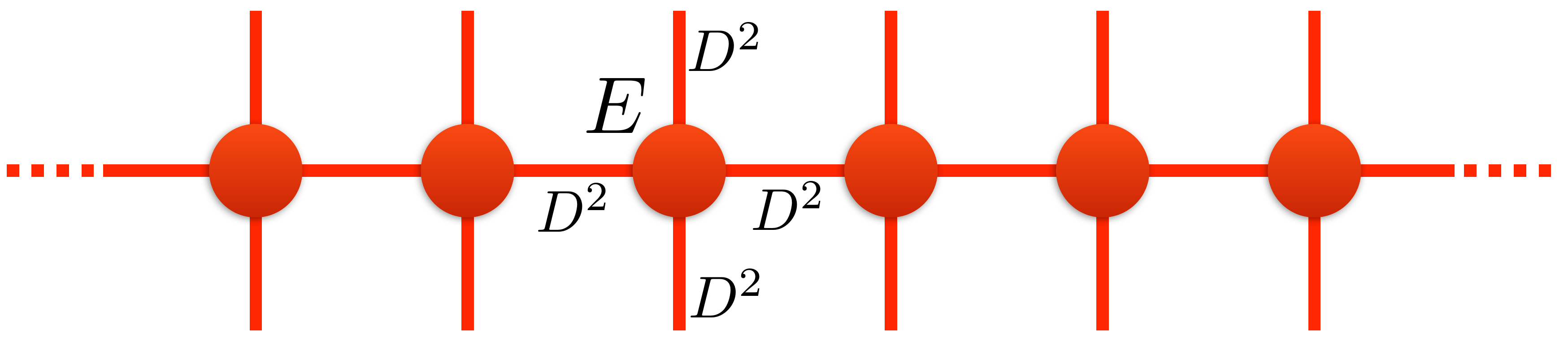}}
	\end{minipage}
	\begin{minipage}[r]{0.17\textwidth}
	\subfloat[]{\includegraphics[width = 12mm, height = 13.5mm]{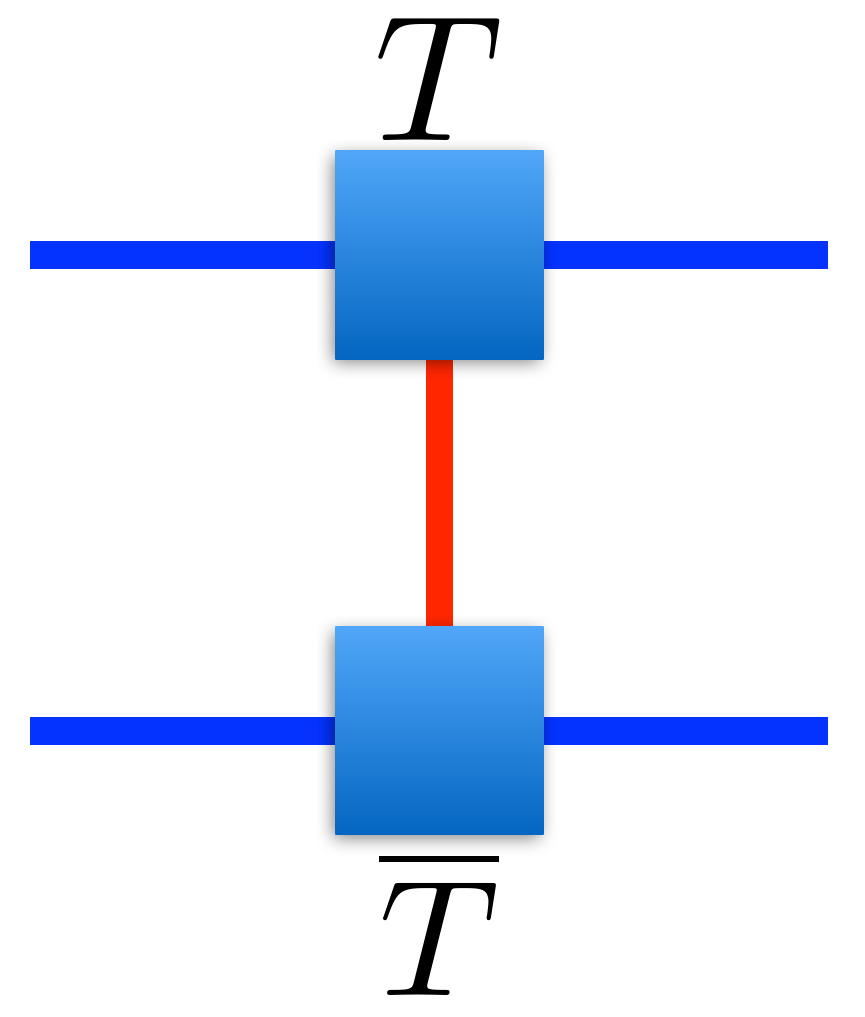}}\\
	\subfloat[]{\includegraphics[width = 28mm, height = 13.5mm]{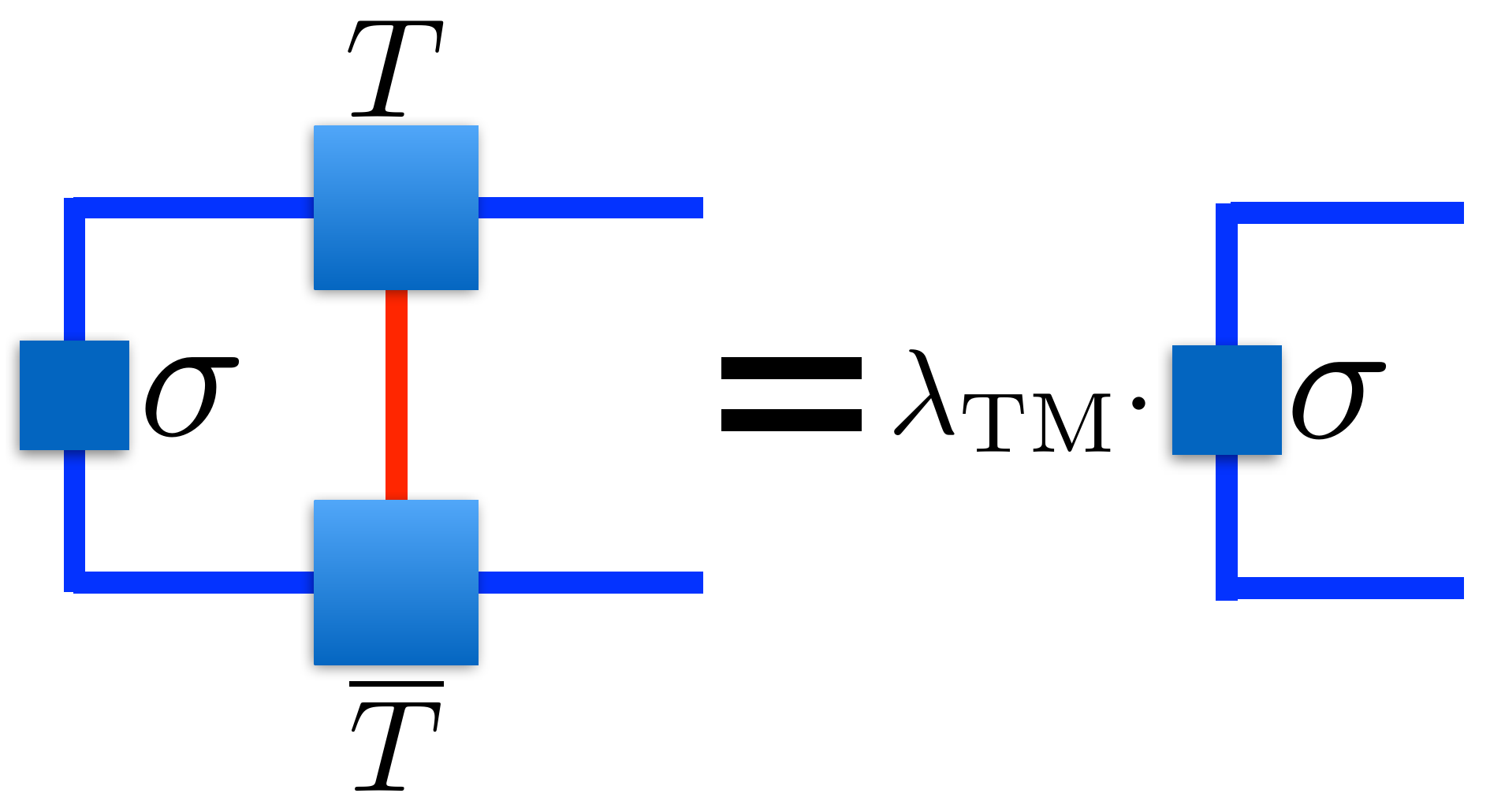}}
	\end{minipage}
\caption{The boundary MPS (a) constructed from the $T$ matrix obtained in CTMRG can be viewed as an approximation of the MPO's (b) leading eigenvector.
%Here the ``physical''  dimension is $D^2$, and the approximation is controlled by MPS's bond dimension $\chi$. 
The transfer matrix of this MPS (c) is reflection symmetric and real, thus has the same left and right leading eigenvector $\sigma$ (d). $\overline{T}$ represents complex conjugate of $T$. 
Entanglement entropy of this MPS can be obtained by analyzing $\sigma$.}
\label{fig:MPS}
\end{figure}

A one-site unit cell MPS defined by the CTMRG $T$ matrix is an approximation of the leading eigenvector of
the matrix product operator (MPO), formed by a line of double tensor $E$, as shown in Fig.~\ref{fig:MPS}(a)(b). Here, the ``physical'' dimension
is $D^2$, and virtual bond dimension is $\chi$, which controls the accuracy of the approximation.
This MPS serves as the boundary state
when contracting the network to obtain physical observables, whose properties inherit from the original PEPS. 

For this boundary MPS, we can extract its correlation length $\xi_{\mathrm{B}}$ and von Neumann entanglement entropy ${\bf S}_{vN}$ of half infinite chain by standard MPS techniques. 
The lattice point group symmetry simplifies the calculation of $\xi_{\mathrm{B}}$ and ${\bf S}_{vN}$. Here $\xi_{\mathrm{B}}$ can be obtained as the largest two eigenvalues of the transfer matrix shown in Fig.~\ref{fig:MPS}(c),
and is in fact very close to $\xi_{\mathrm{max}}^{\mathrm{L,T}}$, since the MPS is the fixed point of the MPO. 
Denoting the leading eigenvector of the transfer matrix as $\sigma$, shown in Fig.~\ref{fig:MPS}(d), after reshaping it into a $\chi\times\chi$ matrix,
the half chain reduced density matrix is given by $\rho=\sigma^2$ up to an isometry. Here proper normalization has been taken.
Then the von Neumann entanglement entropy ${\bf S}_{vN}$ is simply given by
${\bf S}_{vN} = -\mathrm{tr}\{\sigma^2\mathrm{ln}({\sigma^2})\}$. 

For the critical PEPS, the associated MPO is also critical, and finite entanglement scaling would reveal that:
\begin{equation}
{\bf S}_{vN} \sim \frac{c}{6}\mathrm{ln}\xi_{\mathrm{B}},
\end{equation}
where $\xi_{\mathrm{B}}$ and ${\bf S}_{vN}$ are both increasing with $\chi$~\cite{Pollmann2009}. For gapped PEPS, the associated MPO is also gapped, and $\xi_{\mathrm{B}}$ and ${\bf S}_{vN}$ will 
saturate to a constant with increasing $\chi$.

\section{Entanglement spectrum}
\begin{figure}[hb]
\centering	\includegraphics[width = 75mm, height = 55mm]{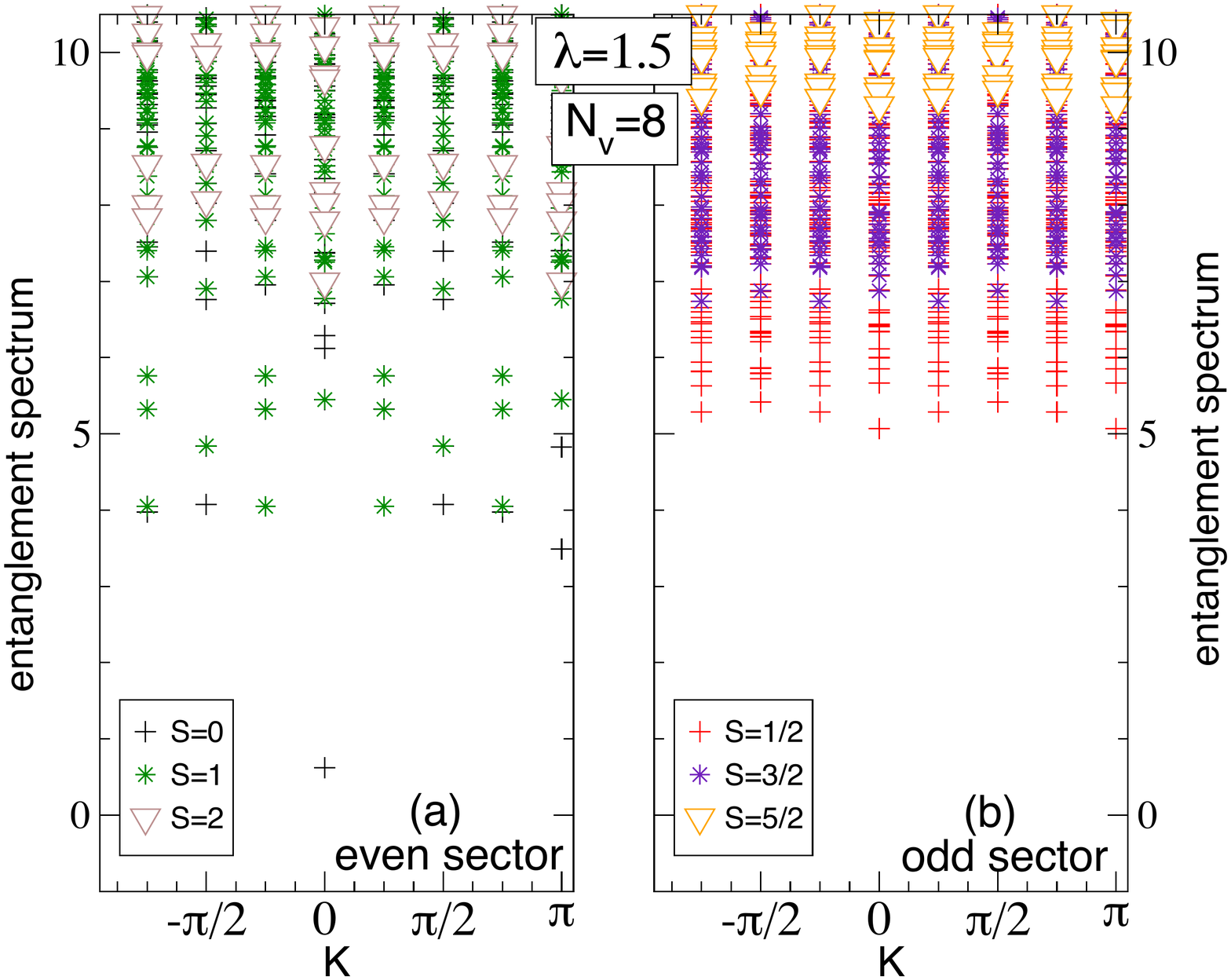}
\caption{Entanglement spectrum at $\lambda=1.5$ computed on an infinite 
cylinder of circumference $N_v=8$. In both even (a) and odd (b) topological sectors, the trace of RDM has been normalized to 1. 
Different symbols are used for the various spin multiplets, according to legends. }
\label{fig:ES}
\end{figure}

For completeness, we have also computed the entanglement spectrum (ES) of the RVB PEPS on a bipartitioned infinite cylinder of
8-site circumference. The latter, defined as the spectrum of minus the logarithm of the reduced density matrix (RDM), can be 
computed exactly from the leading eigenvector of the finite-dimensional transfer matrix (defined in 
Fig.~\ref{fig:MPS}(b) with  periodic boundary conditions) of the (horizontal) cylinder~\cite{Cirac2011}. 
A typical ES of the $\mathbb{Z}_2$ phase is shown in Fig.~\ref{fig:ES} as a function of the momentum $K$ along the vertical cut, 
clearly revealing
the existence of a gap above the ``ground state'', as in the case of the $\lambda=0$ RVB phase~\cite{Poilblanc2012}.
Hence, in contrast to the entanglement entropy of the boundary state, no qualitative difference in the ES 
is seen between the topological and critical phases.  

\begin{figure}[hb]
\centering	\includegraphics[angle=0, width = 55mm, height = 45mm]{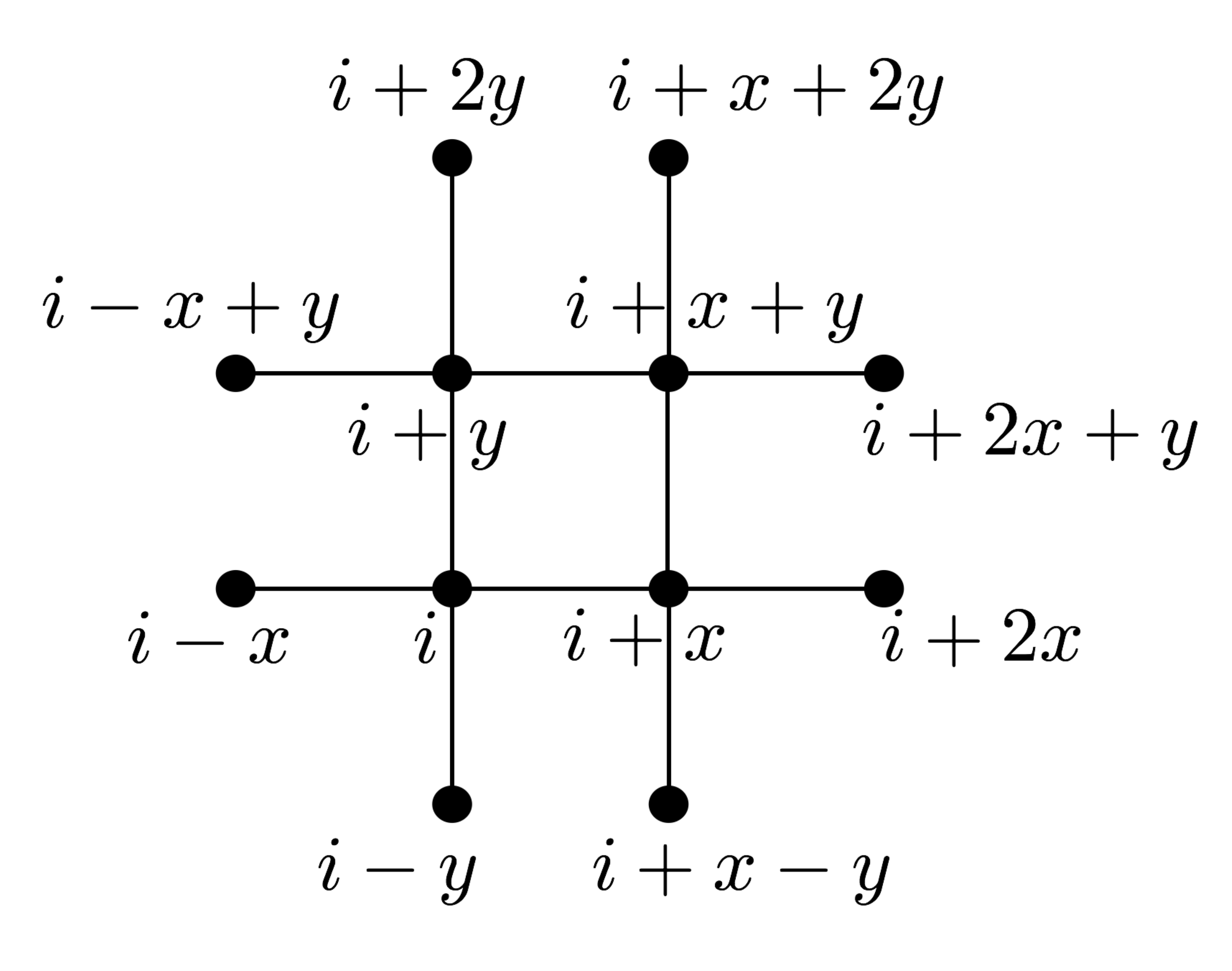}
\caption{Labelling of the sites involved in the definition of the Cano-Fendley Hamiltonian. }
\label{fig:PLAQ}
\end{figure}

\section{Cano-Fendley parent Hamiltonian of the $\lambda=0$ RVB state}

We report here the expression of the $SU(2)$-symmetric parent Hamiltonian of the (critical) RVB state~\cite{Fendley,Mambrini2015} 
defined as a sum over extended 12-site plaquette (see Fig.~\ref{fig:PLAQ}) terms: 
\begin{align}
&H_{\rm CF}(s)=\sum_i \{ K{\cal P}_{i;i-x;i-y;i+x;i+y}^{(\frac{5}{2})}  \nonumber \\
&+{\cal P}_{i-x;i;i-y}^{(\frac{3}{2})} {\cal P}_{i+x;i+y}^{(s)} {\cal P}_{i+x+2y;i+x+y;i+2x+y}^{(\frac{3}{2})} \label{eq:CF} \\
&+{\cal P}_{i+x-y;i+x;i+2x}^{(\frac{3}{2})} {\cal P}_{i;i+x+y}^{(s)} {\cal P}_{i-x+y;i+y;i+2y}^{(\frac{3}{2})} \} \, , \nonumber \\
\nonumber
\end{align}
where $s=0$ or $s=1$ and ${\cal P}_{i_1;i_2;\cdots}^{(S)}$ are projectors on the spin-$S$ manifold within the set of 
(2, 3 or 5) sites $\{i_1;i_2;\cdots\}$. The coefficient $K$ of the Klein term has to be large enough in order to project out non-singlet ground
states. 
Note that the  projectors on $S=0$, $S=1$ or $S=3/2$ ($S=5/2$) can easily 
be expressed as a sum of (product of two) short distance spin-spin interactions~\cite{Mambrini2015}. Therefore, the
resulting Hamiltonian includes only terms like ${\bf S}_i\cdot {\bf S}_j$, $({\bf S}_i\cdot {\bf S}_j)({\bf S}_k\cdot {\bf S}_l)$ or $({\bf S}_i\cdot {\bf S}_j)({\bf S}_k\cdot {\bf S}_l)({\bf S}_n\cdot {\bf S}_m)$, involving at most 6 different sites. 
The coefficients of these various terms, fine-tuned in the CF Hamiltonian of Eq.~\ref{eq:CF}, can be varied to generate a large family of
Hamiltonians which might host the $\mathbb{Z}_2$ SL in some corner of its variational space. 

\nocite{apsrev41Control}
\bibliographystyle{apsrev4-1}
\bibliography{squareZ2SL}